%% file: kappa.tex
\documentclass[aps,prd,twocolumn,superscriptaddress,preprintnumbers,floatfix,nofootinbib,notitlepage,showkeys,showpacs]{revtex4-1}

\usepackage{amsmath,amssymb,amsfonts,mathtools} 
\usepackage[normalem]{ulem}
\usepackage[utf8]{inputenc} 
\usepackage[T1]{fontenc}
\usepackage{hyperref}
\usepackage{graphicx}
\usepackage{latexsym}
\usepackage{bbm}
\usepackage{epstopdf}
\usepackage{color}
\usepackage{braket}
\usepackage{hyperref}
\usepackage{lineno}
\usepackage{slashed}
\usepackage{placeins} 
\usepackage{diagbox}
\usepackage[dvipsnames]{xcolor}

\newcommand{\dd}[1]{\mathop{\mathrm{d}#1}}

\newcommand{\Nc}{N_{\rm c}}
\newcommand{\Ns}{N_{\rm s}}
\newcommand{\Nt}{N_{\rm t}}
\newcommand{\Tc}{T_{\rm c}}
\newcommand{\CF}{C_{\rm F}}

\newcommand{\nn}{\nonumber \\}

\newcommand{\nB}{n_\mathrm{B}}
\newcommand{\GE}{G_\mathrm{E}}
\newcommand{\ZE}{Z_\mathrm{E}}
\newcommand{\tauT}{\tau T}

\newcommand{\GEnorm}{G_\mathrm{E}^\mathrm{norm}}

\renewcommand{\nB}[1]{n_\mathrm{B{#1}}}
\newcommand{\mE}{m_\mathrm{E}}

\newcommand{\ReTr}{\textrm{Re}\,\textrm{Tr}\,}

\newcommand{\be}{\begin{equation}} 
\newcommand{\ee}{\end{equation}}
\newcommand{\bea}{\begin{eqnarray}} 
\newcommand{\eea}{\end{eqnarray}}
\def\lsim{\mathrel{\raise.3ex\hbox{$<$\kern-.75em\lower1ex\hbox{$\sim$}}}}
\def\gsim{\mathrel{\raise.3ex\hbox{$>$\kern-.75em\lower1ex\hbox{$\sim$}}}}

\begin{document}

\title{Lattice QCD constraints on the heavy quark diffusion coefficient}

\author{Nora Brambilla}
\email{nora.brambilla@ph.tum.de}
\affiliation{Physik Department, Technische Universit\"at M\"unchen,
James-Franck-Strasse 1, 85748 Garching, Germany}
\affiliation{Institute for Advanced Study, Technische Universit\"at M\"unchen,
Lichtenbergstrasse 2a, 85748 Garching, Germany}

\author{Viljami Leino}
\email{viljami.leino@tum.de}
\affiliation{Physik Department, Technische Universit\"at M\"unchen,
James-Franck-Strasse 1, 85748 Garching, Germany}

\author{Peter Petreczky}
\email{petreczk@bnl.gov}
\affiliation{Physics Department, Brookhaven National Laboratory,
  Upton, New York 11973, USA}

\author{Antonio Vairo}
\email{antonio.vairo@ph.tum.de}
\affiliation{Physik Department, Technische Universit\"at M\"unchen,
James-Franck-Strasse 1, 85748 Garching, Germany}

\collaboration{TUMQCD Collaboration}
\noaffiliation

\date{\today}
\preprint{TUM-EFT 131/19}

\begin{abstract}
We report progress toward computing the heavy quark momentum diffusion coefficient from the correlator
of two chromoelectric fields attached to a Polyakov loop in pure SU(3) gauge theory.
Using a multilevel algorithm and tree-level improvement, we study the behavior of the diffusion coefficient as a function
of temperature in the wide range $1.1<T/\Tc<10^4$
in order to compare it to perturbative expansions at high temperature.
We find that within errors the lattice results are remarkably compatible with the next-to-leading order
perturbative result.
\end{abstract}

\maketitle

\section{Introduction}\label{sec:intro}

The matter produced in heavy ion collisions can be described as a nearly ideal fluid;
see Ref.~\cite{Busza:2018rrf} for a recent review.
Because of the high energy density, the created matter is deconfined and can be characterized as
a strongly coupled quark-gluon plasma (sQGP)~\cite{Shuryak:2003ty,Shuryak:2003xe}.
One recently realized interesting feature of the quark gluon plasma 
is the fact that heavy quarks participate in the collective behavior,  
see Ref.~\cite{Rapp:2018qla} for a recent review.
This is interesting for the following reason:
The relaxation time of heavy
quarks is expected to be $\sim (M/T) t_{\rm rel}^{\rm light}$, with $M$ being the heavy quark mass,
$T$ being the temperature, and $t_{\rm rel}^{\rm light}$ being the relaxation time of the bulk (light)
degrees of freedom in sQGP. 
The lifetime of the hot medium created in heavy ion collisions is about $5-10$ fm. 
Since the collectivity in the heavy quark sector implies 
that the relaxation time of the heavy quark is much shorter 
than the lifetime of the medium despite the enhancement factor of $M/T$, 
this in turn means that the relaxation time of the bulk degrees of freedom is very short,
thus further corroborating the strongly coupled nature of the matter produced in heavy ion collisions.

Because the relaxation time of heavy quarks is much larger than the relaxation time of light
degrees of freedom, the dynamics of heavy quarks can be understood in terms of Langevin equations~\cite{Moore:2004tg}.
The drag coefficient $\eta$ and the heavy quark momentum diffusion coefficient $\kappa$ 
that enter into the Langevin equations describe the interaction of the heavy quarks with the medium
and are connected by the Einstein relation $\eta=\kappa/(2MT)$ in thermal equilibrium. 
The heavy quark diffusion coefficient
has been calculated in perturbation theory at leading order (LO)~\cite{Moore:2004tg,Svetitsky:1987gq},
as well as at next-to-leading order (NLO)~\cite{CaronHuot:2008uh}. The NLO correction is very large,
thus calling into question the validity of the perturbative expansion. Analytic calculations for
strong coupling are available only for supersymmetric Yang-Mills theories~\cite{Herzog:2006gh,CasalderreySolana:2006rq}.
Therefore, lattice QCD calculations for the heavy quark diffusion coefficient are needed. 

It is well known, however, that lattice calculations of the transport coefficients are very difficult. 
To obtain the transport coefficients one has to reconstruct the spectral functions from the appropriate
Euclidean time correlation functions. 
At low energies, $\omega$, the spectral function has a peak, called the transport peak, 
and the width of the transport peak defines the transport coefficient.
Thus, one needs a reliable determination of the width of the transport peak in order to obtain
the transport coefficient from lattice QCD calculations, which is difficult~\cite{Aarts:2002cc,Petreczky:2005nh}. 
In the case of heavy quarks, this is even more challenging 
because the width of the transport peak is inversely proportional to the heavy quark mass. 
Moreover, Euclidean time correlators 
are rather insensitive to small widths~\cite{Petreczky:2005nh,Petreczky:2008px,Ding:2012sp,Ding:2018uhl,Lorenz:2020uik,Borsanyi:2014vka}.
Recently the problem of heavy quark diffusion has also been studied out of equilibrium 
with real-time lattice simulations in Refs.~\cite{Boguslavski:2020mzh,Boguslavski:2020tqz}.
Moreover, the heavy quark momentum diffusion coefficient is a crucial parameter entering the 
evolution equations describing the out-of-equilibrium dynamics 
of heavy quarkonium in sQGP~\cite{Brambilla:2016wgg,Brambilla:2017zei,Brambilla:2019tpt}.

The above difficulty in the determination of the heavy quark diffusion coefficient can be circumvented by
using an effective field theory approach. Namely, by integrating out the heavy quark fields, one can relate the heavy quark
diffusion coefficient to the correlator of the chromoelectric field strength~\cite{CaronHuot:2009uh}.
The corresponding spectral function does not have a transport peak, and the small $\omega$ behavior is smoothly
connected to the UV behavior of the spectral function~\cite{CaronHuot:2009uh}. The heavy quark diffusion coefficient
is given by the intercept of the spectral function at $\omega=0$ and no determination of the width of the transport peak
is needed. Lattice calculations of $\kappa$ along these lines have been carried out in the SU(3) gauge theory in the
deconfined phase i.e., for purely gluonic plasma~\cite{Meyer:2010tt,Francis:2011gc,Banerjee:2011ra,Francis:2015daa,Shu:2019twy}.
The correlator of the chromoelectric field strength is very noisy, making the lattice calculations extremely
challenging. To deal with this problem, it is mandatory to use noise-reducing techniques such as the
multilevel algorithm by L\"uscher and Weisz~\cite{Luscher:2001up}. This algorithm is based on the locality of the action
and therefore is only available for the pure gauge theory. This is the reason why the calculations of the heavy quark
diffusion coefficient are performed in the SU(3) gauge theory. Another challenge in the determination of the heavy quark
diffusion coefficient is the reconstruction of the spectral function from 
the Euclidean time correlation function. 
The above lattice studies used a simple parameterization of the spectral function to extract $\kappa$. 
One has to explore the sensitivity of the results on the parameterization of the spectral function. 
More generally, one has to understand to 
what extent the Euclidean time correlation function of the chromoelectric field strength is sensitive 
to the small $\omega$ behavior of the corresponding spectral function. 

At sufficiently high temperatures the perturbative calculations of the heavy quark diffusion coefficient should
be adequate. This suggests that $\kappa/T^3$ should decrease from large values at temperatures close to
the transition temperature to smaller values when the temperature is increasing. It would be interesting to see
if contacts between the lattice and the perturbative calculations can be made for the heavy quark diffusion
coefficient, as has already been done for the equation of state~\cite{Bazavov:2017dsy}, 
quark number susceptibilities~\cite{Bazavov:2013uja,Ding:2015fca},
and static correlation functions~\cite{Bazavov:2016uvm,Bazavov:2018wmo,Bazavov:2019www}.
If such contacts can be established, these would validate the methodology used in the lattice extraction of $\kappa$.
Previous lattice studies focused on a narrow temperature region~\cite{Banerjee:2011ra}
or only considered a single value of the temperature~\cite{Francis:2015daa}. 
In Ref.~\cite{Banerjee:2011ra}, no significant temperature dependence of $\kappa/T^3$ was found.
Large temperatures are needed in the lattice studies to establish the temperature dependence of $\kappa/T^3$.
The temperature dependence of $\kappa/T^3$ is also important for phenomenology, as with a constant value
of $\kappa/T^3$ it is impossible to explain simultaneously the elliptic flow parameter, $v_2$, for heavy quarks and
the nuclear modification factor~\cite{Rapp:2018qla}. 
Furthermore, the spectral function of the chromoelectric field strength correlator is known at NLO~\cite{Burnier:2010rp}.
Using this NLO result at high $\omega$ one can constrain the functional form of the spectral function used in the analysis
of the lattice correlator. 

The aim of this paper is to study the correlator of the chromoelectric field strength in a wide temperature range 
in order to make contact with weak coupling calculations of the Euclidean correlation function up to NLO in the spectral function,
and also to constrain the temperature dependence of $\kappa$.

The rest of the paper is organized as follows: In the next section,
we go through the procedure of calculating the Euclidean correlator of the chromoelectric field strength on the lattice. 
The spectral function of the chromoelectric correlator and its relation to $\kappa$ is discussed 
in Sec~\ref{sec:rho}. There we also review the perturbative results for this spectral function.
The short-time behavior of the chromoelectric correlator and its proper normalization
is clarified in Sec~\ref{sec:cont}. In Sec~\ref{sec:model}, we discuss how to model the spectral functions
of the chromoelectric correlator and to extract the value of $\kappa$ from the lattice results. 
Finally, Sec~\ref{sec:concl} contains our conclusions.

\section{Lattice results for the chromoelectric correlator}\label{sec:meas}

For a heavy quark of mass $M\gg\pi T$, 
the heavy quark effective theory (HQEFT) provides 
a method of calculating the heavy quark diffusion coefficient in the heavy quark limit
by relating it to a chromoelectric correlator
in Euclidean time~\cite{CaronHuot:2009uh,CasalderreySolana:2006rq}:
\be\label{eq:gelat}
\GE(\tau) = -\sum_{i=1}^{3} 
 \frac{\left\langle \ReTr\left[U(1/T,\tau)E_i(\tau,\mathbf{0})U(\tau,0)E_i(0,\mathbf{0})\right]\right\rangle}{3\left\langle\ReTr U(1/T,0)\right\rangle}\,,
\ee
where $T$ is the temperature, $U(\tau_1,\tau_2)$ is the temporal Wilson line between $\tau_1$ and $\tau_2$,
and the chromoelectric field, 
in which the coupling has been absorbed $E_i\equiv gE_i$, is discretized on the lattice as~\cite{CaronHuot:2009uh}:
\be\label{eq:elfield}
E_i(\mathbf{x},\tau) = U_i(\mathbf{x},\tau) U_4(\mathbf{x}+\hat{i},\tau) - U_4(\mathbf{x},\tau)U_i(\mathbf{x}+\hat{4})\,.
\ee
This discretization is expected to be the least sensitive to ultraviolet effects~\cite{CaronHuot:2009uh}.

To calculate the discretized chromoelectric correlator defined above on the lattice, we use the standard Wilson gauge
action and the multilevel algorithm~\cite{Luscher:2001up}. We consider $\Ns^3 \times \Nt$ lattices and vary the 
temperature in a wide range $T=1.1\,\Tc - 10^4\,\Tc$ by varying the lattice gauge coupling $\beta=6/g_0^2$.
Here $\Tc$ is the deconfinement phase transition temperature.
We use $\Nt=12,~16,~20$, and $24$ at each temperature to check for lattice spacing effects
and perform the continuum extrapolation. In this study we use $\Ns=48$, except for $\Nt=12$ lattices,
where multiple spatial volumes are used to check for finite-volume effects.

To set the temperature scale as well as the lattice spacing, we use the gradient flow parameter 
$t_0$~\cite{Luscher:2010iy} and the value $\Tc \sqrt{t_0}=0.2489(14)$~\cite{Francis:2015lha}. We use
the result of Ref.~\cite{Francis:2015lha} to relate the temperature scale or the lattice spacing
to $\beta$. The parameters of the lattice calculations, including the statistics,
are given in Table~\ref{tab:confs}. In the simulations with the multilevel algorithm,
we divide the lattice into four sublattices and update each sublattice 2000 times
to evaluate the chromoelectric correlator on  a single gauge configuration.
We use the simulation program developed in a prior study~\cite{Banerjee:2011ra}.

In order to obtain the heavy quark diffusion coefficient,
the lattice chromoelectric correlator needs to be renormalized and then extrapolated to the continuum.
\footnote{
We will use the notation $\GE$ for both the lattice and 
the continuum version of the chromoelectric correlator to keep the notation simple. 
It should be clear from the context which one we are referring to.
We will use different notations for the continuum and the lattice version of the chromoelectric correlator 
only when it is absolutely necessary.}
The renormalization coefficient $\ZE(\beta) \equiv \ZE(g_0^2)$ of the chromoelectric correlator
in the case of the Wilson gauge action has been calculated at one loop~\cite{Christensen:2016wdo}:
\be\label{eq:pertre}
\ZE^{\mathrm{1-loop}} = 1+0.1377185690942757(4)g_0^2\,.
\ee
We will use this one loop correction in the present study. 
However, we expect that the one loop result for $\ZE$ is not precise enough. 
As will be clear from the results of the lattice calculations, 
this is indeed the case. The perturbative error in $\ZE(\beta)$ affects both
its absolute value for fixed $\beta$ and its $\beta$-dependence. For the continuum extrapolation,
it is important to estimate the uncertainty in the $\beta$-dependence of the renormalization constant.
The error in the absolute value of $\ZE$ could be corrected after the continuum extrapolation is done
by introducing an additional multiplicative factor. 
We will postpone the discussion of this multiplicative factor to Sec~\ref{sec:cont}.
To estimate the error in the $\beta$ dependence  of
$\ZE$ we consider the tadpole improved result for $\ZE$, namely $\ZE^{\mathrm{tad}}=1/u_0$, with $u_0$ being the
plaquette expectation value~\cite{Banerjee:2011ra}. 
The difference in the $\beta$ dependence of $\ZE^{\mathrm{tad}}$ and $\ZE^{\mathrm{1-loop}}$ 
can be used as an estimate of the error
of the $\beta$-dependence of $\ZE$. Therefore, at each temperature we consider the variation
in $\ZE^{\mathrm{1-loop}} \cdot u_0$ in the $\beta$ range that corresponds to $\Nt=12-24$ as an estimate
of the systematic errors in $\ZE$ for bare gauge couplings in that range.
\begin{table}
  \begin{ruledtabular}
    \begin{tabular}{c|ccc}
      \input{table_Nconfs.dat}
    \end{tabular}
  \end{ruledtabular}
  \caption{Parameters of the lattice calculations.}
  \label{tab:confs}
\end{table}%

\begin{figure}[!ht]
  \includegraphics[width=8.6cm]{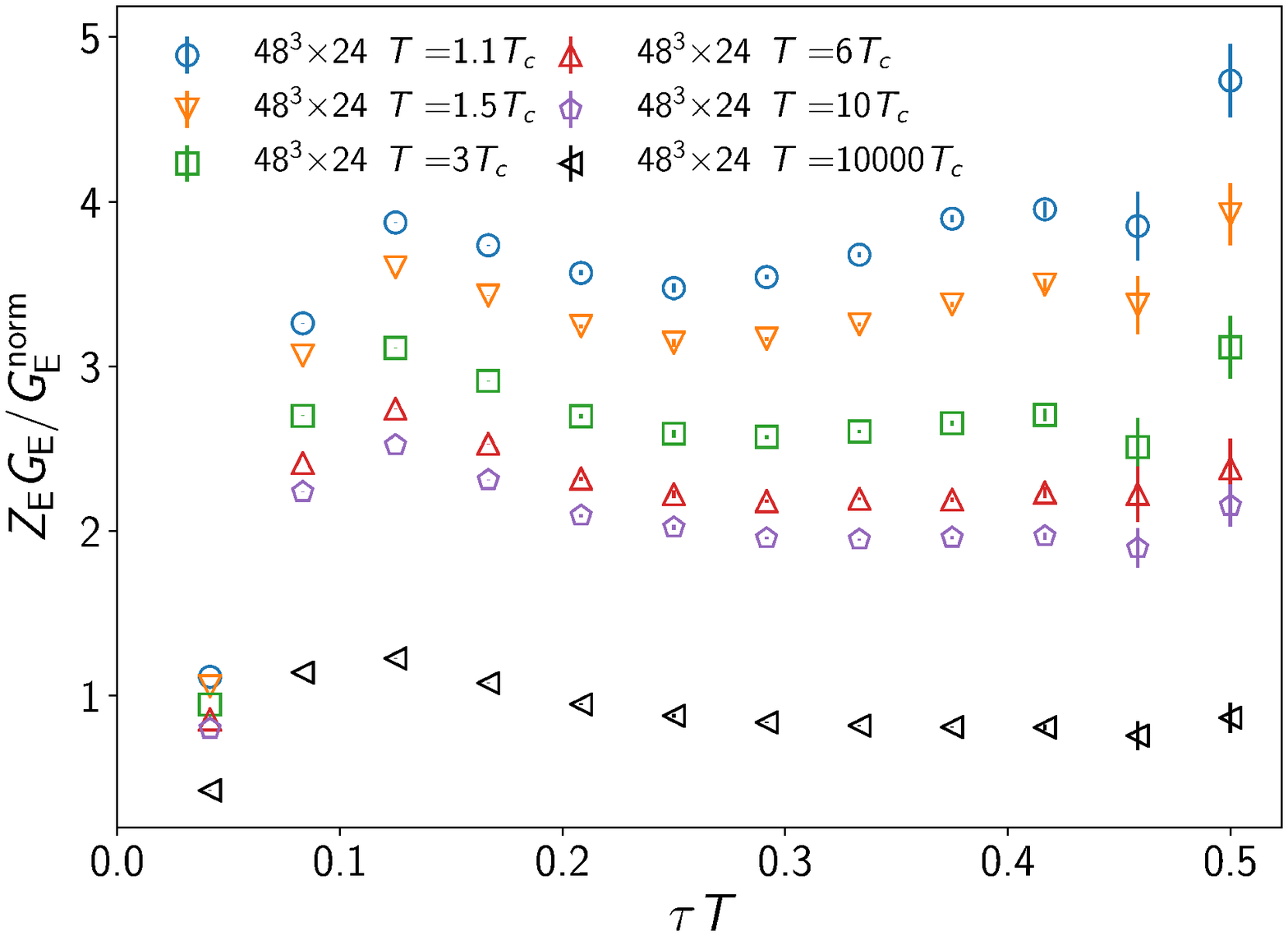}
  \includegraphics[width=8.6cm]{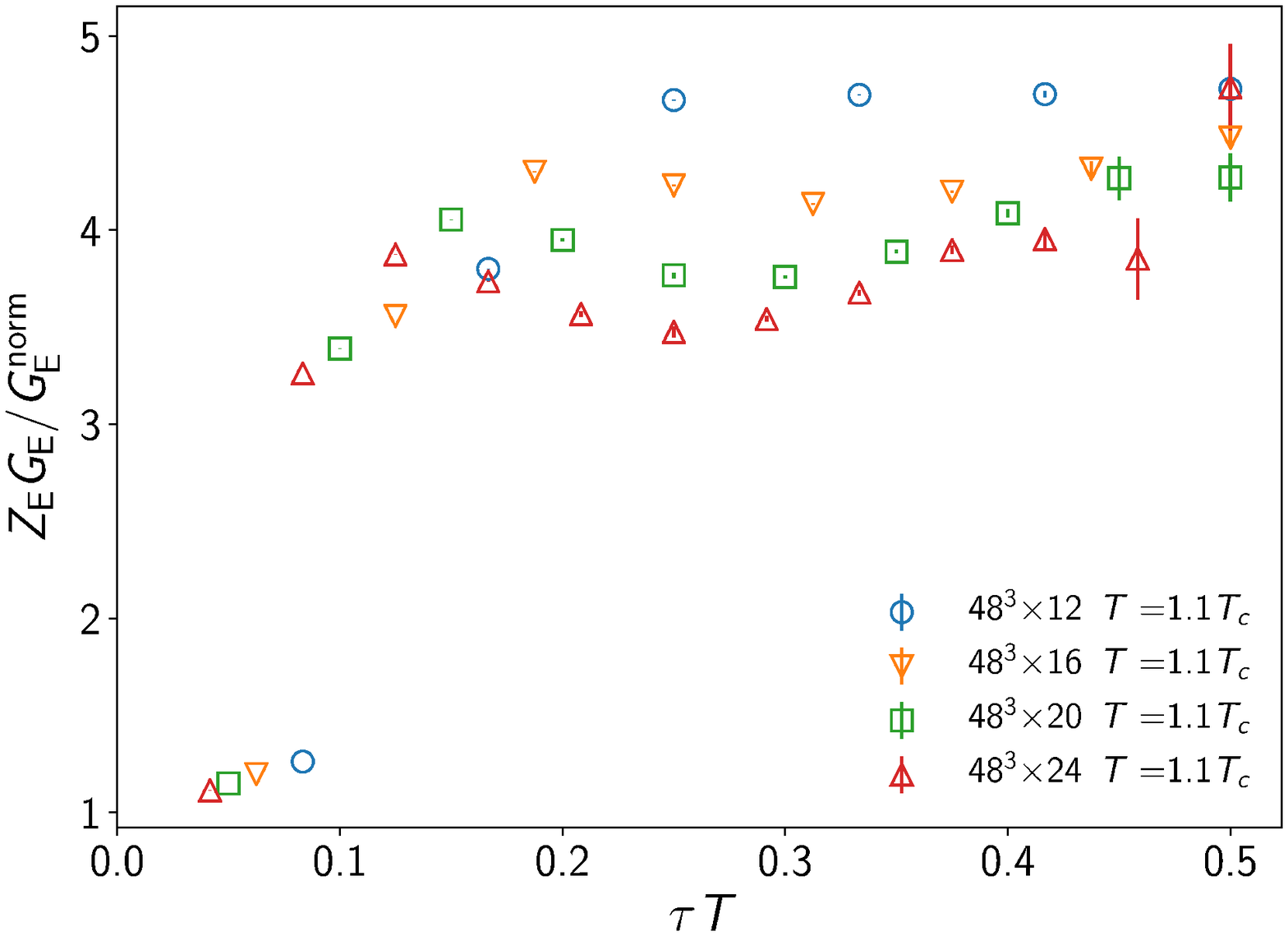}
  \caption[b]{The chromoelectric field correlator from Eq.~\eqref{eq:gelat} normalized with Eq.~\eqref{eq:gepert}.  
              Top: all measured temperatures for $\Nt = 24$.
			  Bottom: all measured temporal extents $\Nt$ for the smallest temperature.
  }
  \label{fig:halfraw} 
\end{figure}

The chromoelectric correlator decays rapidly with increasing $\tau$. This feature can be understood from
the leading order (tree-level) result~\cite{CaronHuot:2009uh}:
\be\label{eq:gepert}
\frac{\GE^\mathrm{LO}(\tau)}{g^2\CF} \equiv \GEnorm(\tau) = \pi^2 T^4 \left[\frac{\cos^2(\pi\tauT)}{\sin^4(\pi \tauT)}+\frac{1}{3\sin^2(\pi\tauT)}\right]\,,
\ee
where $\CF=4/3$ is the Casimir of the fundamental representation of SU(3).
In Fig.~\ref{fig:halfraw} we show $\ZE \GE/\GEnorm$ for different temperatures calculated on the largest, 
$48^3 \times 24$ lattice. We see a significant temperature dependence in this ratio.
Also shown in the figure are the numerical results for 
the lowest temperature, $T=1.1\,\Tc$ calculated for different $\Nt$.  
As one can see from the figure, the cutoff ($\Nt$) dependence is significant even for relatively large values of $\tauT$.  
We expect that the cutoff dependence increases with decreasing $\tauT$, except when $\tau$ is 
of the order of the lattice spacing because the cutoff dependence of $\ZE \GE/\GEnorm$ is proportional to $(a/\tau)^2$.
We see that our lattice data follow this expectation for $\tauT>0.2$.
This observation is important for estimating the reliability of the continuum extrapolations.
A similar $\Nt$ dependence is observed at other temperatures.

\begin{figure}[!ht]
  \includegraphics[width=8.6cm]{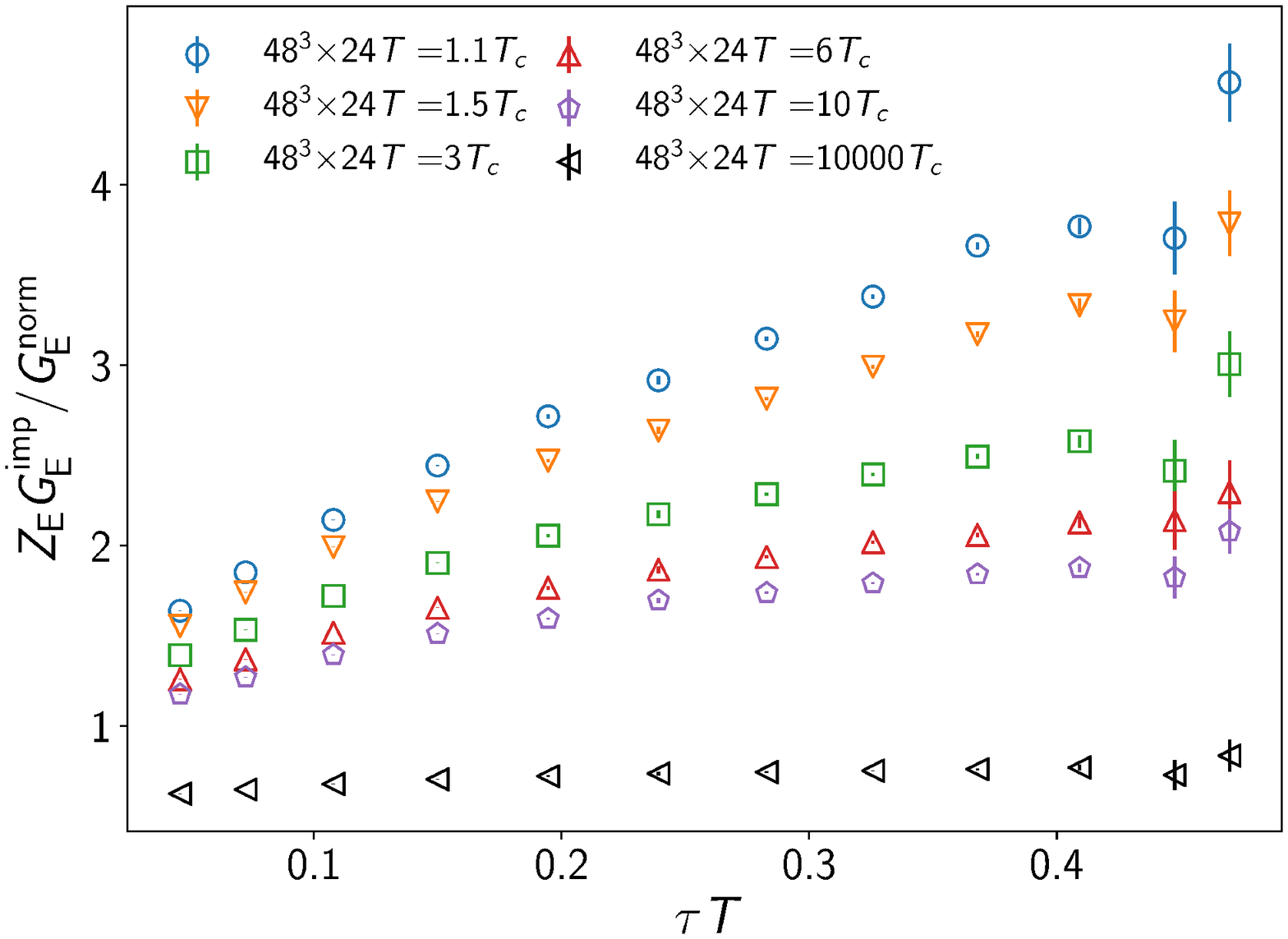}
  \includegraphics[width=8.6cm]{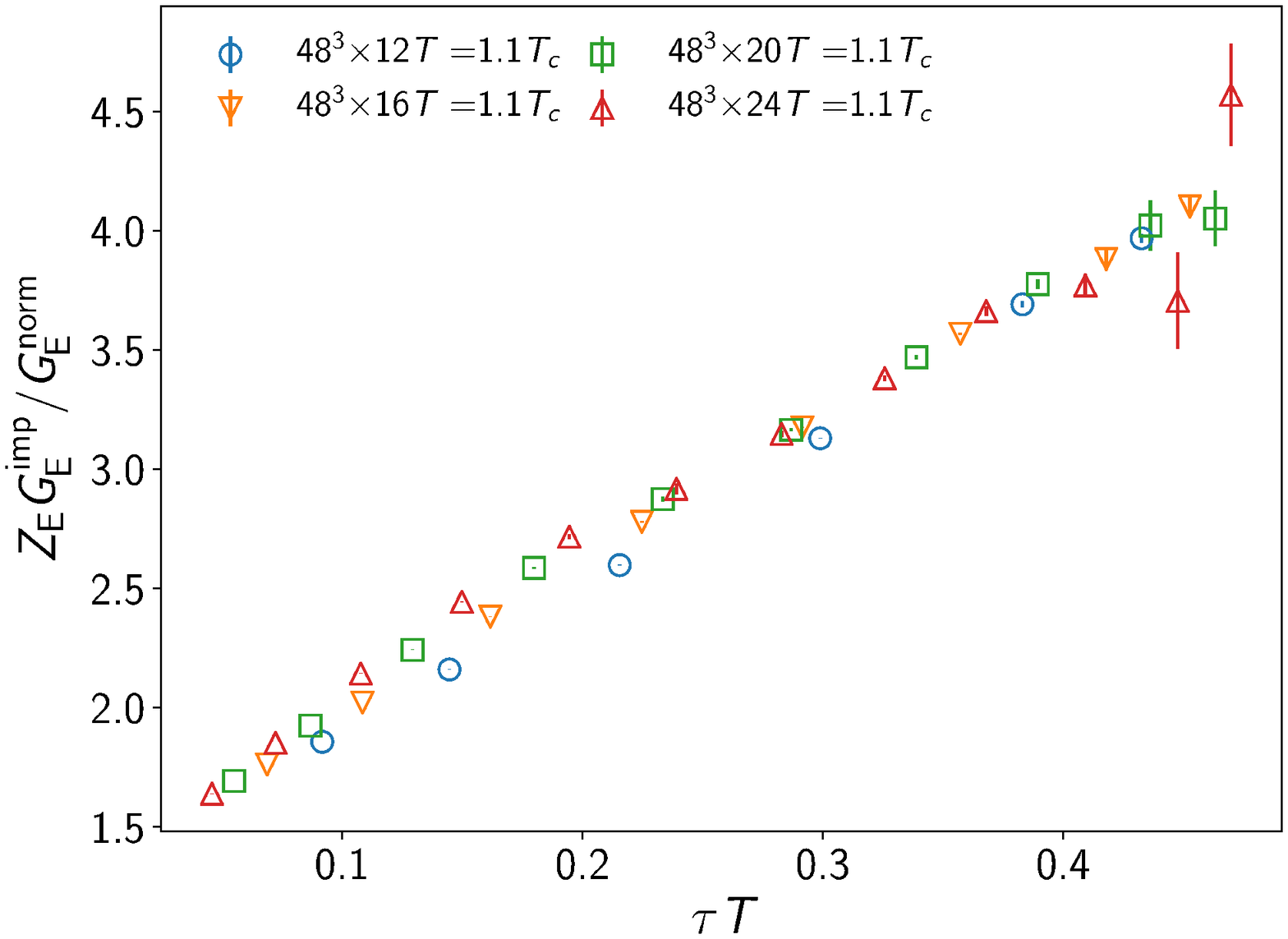}
  \caption[b]{The chromoelectric field correlator from Eq.~\eqref{eq:gelat} normalized with Eq.~\eqref{eq:gepert} 
              and tree-level improved with~\eqref{eq:lolatpert}.
              Top: all measured temperatures for the biggest $\Nt = 24$. 
              Bottom: all measured temporal extents $\Nt$ for the smallest temperature.
  }
  \label{fig:notraw} 
\end{figure}

In order to reduce discretization errors we turn to a tree-level improvement procedure~\cite{Sommer:1993ce,Meyer:2009vj},
where the leading order results in the continuum~\eqref{eq:gepert} and the lattice perturbation theory are matched.
The LO lattice perturbation theory gives~\cite{Francis:2011gc}:
\begin{align}\label{eq:lolatpert}
\frac{\GE^\mathrm{LO,lat}(\tau)}{g^2\CF} &= 
\int_{-\pi}^{\pi} \frac{\mathop{\mathrm{d}^3 q}}{(2\pi)^3}
\frac{\tilde{q}^2e^{\bar{q}\Nt(1-\tauT)} + \tilde{q}^2e^{\bar{q}\Nt \tauT}}{3a^4\left(e^{\bar{q}\Nt}-1\right)\sinh(\bar{q})}\,,\\
\intertext{where}
\bar{q} &= 2\mathrm{arsinh}(\tilde{q}/2)\,, \\
\tilde{q}^2 &= \sum_{i=1}^3 4\sin^2(q_i/2)\,.
\end{align}
The improved distance $\overline{\tau}$ is then defined so that $\GE^\mathrm{LO}(\overline{\tau}) = \GE^\mathrm{LO,lat}(\tau)$.
In Fig.~\ref{fig:notraw}, we show our results for $\ZE \GE^\mathrm{imp}(\tau)/\GEnorm = \ZE \GE(\overline{\tau})/\GEnorm$.
From the figure, we can observe that after the tree-level improvement, the ratio $\ZE \GE/\GEnorm$ 
appears monotonically increasing with increasing $\tauT$ and has a decreasing slope as a function of temperature. 
At the highest temperature, $T=10^4\,\Tc$, we see a nearly horizontal $\tau$-independent line.
Moreover, we observe a large reduction of cutoff effects for all temperatures when tree-level improvement is used. 
As an example, we show this reduction at the bottom of Fig.~\ref{fig:notraw} for the lowest temperature, $T=1.1\,\Tc$.
A similar reduction in the $\Nt$ dependence is seen at other temperatures.
Due to its impact, we will use the tree level improvement for the rest of this paper, 
and therefore, unless otherwise indicated, drop the overline from $\overline{\tau}$ and the superscript "imp" from $\GE^\mathrm{imp}$.
\begin{figure}[!ht]
  \includegraphics[width=8.6cm]{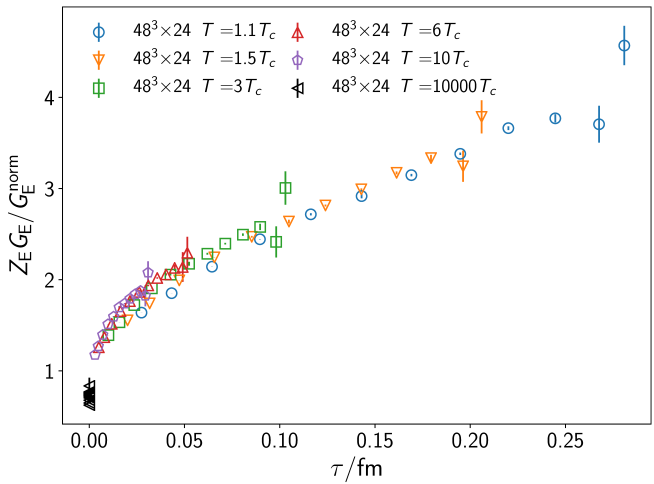}
  \caption[b]{The data of Fig~\ref{fig:notraw} in physical units.
  }
  \label{fig:fermipert} 
\end{figure}%

\begin{figure}[!ht]
  \includegraphics[width=8.6cm]{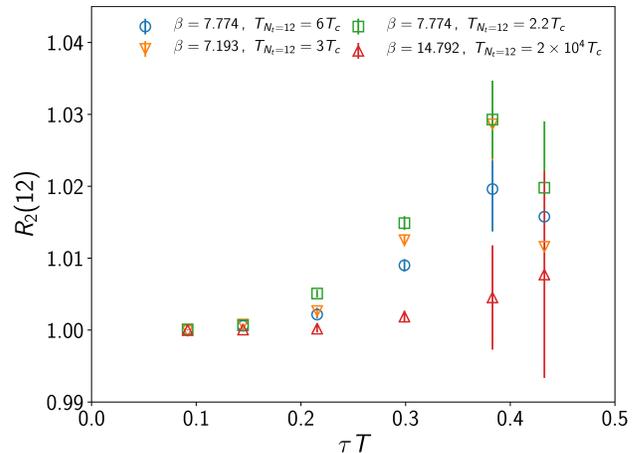}
  \caption[b]{The ratio Eq.~\eqref{eq:ratiodef} of simulations with the same $\beta$ at different temperatures with $\Nt=12$.
  }
  \label{fig:therratio} 
\end{figure}

The normalized chromoelectric correlator shown in Fig.~\ref{fig:notraw} 
has a significant $\tau$-dependence. We conclude that the LO
perturbative result does not capture the key features of the chromoelectric correlator. 
Only at the highest temperature, $T = 10^4\,\Tc$, 
is the $\tau$-dependence of the correlator well described by the leading order result.
One may wonder whether the observed behavior of the normalized chromoelectric correlator is due to thermal effects 
that are not present at leading order, 
like the physics of the heavy quark transport, or are due to higher-order effects at zero temperature. 
In order to answer this question, we show our lattice results in Fig.~\ref{fig:fermipert} 
as a function of $\tau$ in physical units rather than a function of $\tauT$.
This figure shows that the ratio of the chromoelectric correlator to the free theory result is largely temperature independent
implying that the chromoelectric correlator is dominated by the vacuum part of the spectral function.
It thus becomes even more important to quantify the temperature dependence of the chromoelectric correlator.
This can be done by considering the
following ratio of the normalized correlator at a fixed value of $\beta$, but at two temperatures
corresponding to temporal extents $\Nt$ and $2 \Nt$. Lattice artifacts are canceled out in the double ratio:
\be\label{eq:ratiodef}
R_2(\Nt) = \frac{\GE(\Nt,\beta)}{\GEnorm(\Nt)}\bigg/\frac{\GE(2\Nt,\beta)}{\GEnorm(2\Nt)}\,.
\ee
Furthermore, if the LO result is a good approximation of the correlator and 
$\kappa/T^3$ is temperature independent, $R_2(\Nt)$ should be 1 and
independent of the temperature, while the temperature dependence of $\kappa/T^3$
will make this ratio different from 1 and also temperature dependent.
The amount by which $R_2(\Nt)$ deviates from 1 also depends on the value of $\kappa/T^3$:
small values of $\kappa/T^3$ will result only in small deviations of $R_2(\Nt)$ from 1.
Our results for $R_2(\Nt)$ are shown in Fig.~\ref{fig:therratio}.
At the highest temperature the double ratio is consistent with 1 within errors,
perhaps not surprisingly as at high temperatures the temperature dependence of $\kappa/T^3$ is expected to be logarithmic, and thus rather mild.
At lower temperatures, however,  we see deviations from 1~in the double ratio 
at the few-percent level, which increase with decreasing temperature and increasing $\tauT$.
On the other hand, for $\tauT<0.2$ the double ratio is close to 1,
implying that there the correlator is dominated by the $T=0$ part of the spectral function.
In any case, thermal effects in the chromoelectric correlator, which encode the value of $\kappa$,
are small, at the level of few percent. This fact implies that extracting
$\kappa$ from lattice determinations of the chromoelectric correlator is challenging.

Before extracting the heavy quark diffusion coefficient we need to address finite-volume 
effects and perform the continuum extrapolation of $\ZE \GE$. 
Most of our calculations have been performed using $\Ns=48$.
To check for finite-volume effects for $\Nt=12$, we have performed calculations
using  spatial sizes $N^3_\mathrm{s}=24^3,32^3,48^3$ at two temperatures, $T=1.5\,\Tc$ and $T=10\,\Tc$.
The smallest spatial volume here corresponds to the aspect ratio $\Ns/\Nt=2$. 
The detailed study of finite-volume effects is discussed in Appendix~\ref{sec:appA}. 
We find that the finite-volume effects are small, considerably smaller than other sources of error down
to the aspect ratio $\Ns/\Nt=2$. Therefore, at the current level of precision, using a $\Ns=48$ lattice
is sufficient even for $\Nt=24$.

\begin{figure}[!ht]
  \includegraphics[width=8.6cm]{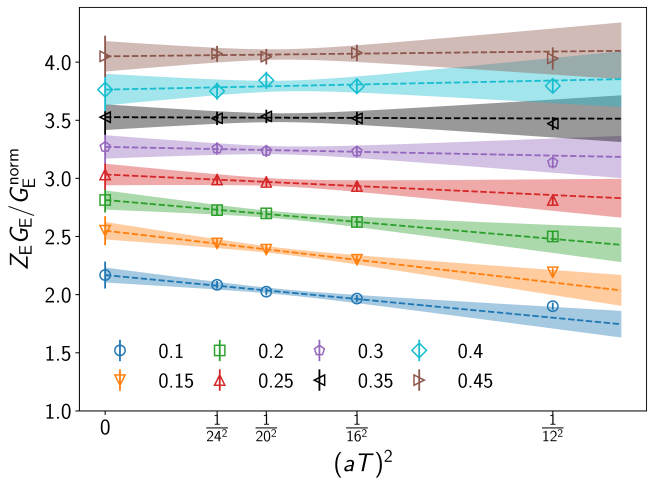}
  \caption[b]{The continuum extrapolation. 
              The lines represent the fit performed with the three largest $\Nt$'s at temperature $T=1.1\,\Tc$
              at different values of $\tauT$, shown with different colors.
              The fitted line is extrapolated to 0 and to $1/12^2$ to show the quality of the fit 
              compared to points at those locations. 
              The point at zero includes the systematic error coming from the inclusion of the smallest lattice.
  }
  \label{fig:contextrap}
\end{figure}

Next, we perform the continuum extrapolations of $\ZE \GE$.
The systematic errors in the renormalization constant estimated above are combined
with the statistical errors of the chromoelectric correlator before performing the
continuum extrapolation. In the interval $0.1 \le \tauT \le 0.45$ we have a sufficient number
of data points to perform the continuum extrapolations. We first interpolate the data
for each $\Nt$ in $\tauT$ using ninth-order polynomials to estimate $\ZE \GE$ at 
common $\tauT$ values. We perform linear extrapolations in $1/\Nt^2=(a T)^2$ of $\ZE \GE$ at
these $\tauT$ values using lattices with $\Nt=16,~20$, and $24$. 
As an example, we show the continuum extrapolation for selected values of $\tauT$ in Fig.~\ref{fig:contextrap}.
One can see 
that the $\Nt=12$ data do not lie in the $1/\Nt^2$ scaling
region. Therefore, we also perform extrapolations to $\Nt=12$ data with a $(a T)^4$ term included.
The difference between these continuum extrapolations is used as an estimate of
the systematic error of the continuum result. The slope of the $a^2$ dependence is increasing
with decreasing $\tauT$, as can be seen from Fig.~\ref{fig:contextrap}. 
This is expected; the cutoff effects are larger at smaller $\tauT$. 
However, at the smallest value, $\tauT=0.1$, the slope
of the $a^2$ dependence becomes smaller again contrary to expectations. 
We take this as an indication that the cutoff effects in this region cannot be
described by a simple $a^2$ or $a^2 + a^4$.  
As shown in Appendix~\ref{sec:appA} the slope of the $a^2$ dependence increases monotonically only till $\tauT \ge 0.175$. 
Therefore, we consider the continuum extrapolation to be reliable only for $\tauT \ge 0.175$. 
For an additional cross-check, we also perform the continuum extrapolation of the lattice data without tree-level improvement. 
This is discussed in Appendix~\ref{sec:appA}, where further details of the continuum extrapolations can be found.  

\begin{figure}[!ht]
  \includegraphics[width=8.6cm]{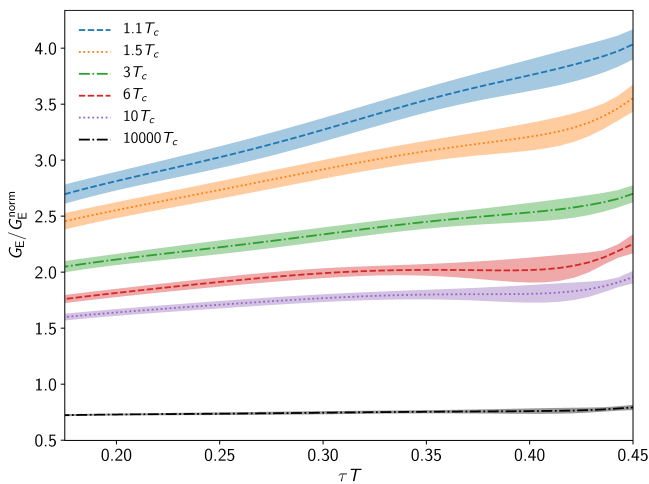}
  \caption[b]{The continuum extrapolation for all temperatures as a function of $\tauT$.
  }
  \label{fig:tempnonorm}
\end{figure}%

The continuum-extrapolated chromoelectric correlator normalized by $\GEnorm$
is shown in Fig.~\ref{fig:tempnonorm} for all temperatures as  function of $\tauT$. 
The continuum-extrapolated results share the general features of the tree-level-improved results at nonzero
lattice spacing in terms of $\tau$ and temperature dependence. 
In particular, we see a strong dependence on $\tauT$, 
except for the highest temperature, indicating that the leading-order result does not capture
the $\tauT$ dependence of $\GE$. 
We will try to understand these features of the correlator in the next sections.

\section{Spectral functions and diffusion coefficient in perturbation theory}\label{sec:rho}

In order to determine the heavy quark diffusion coefficient $\kappa$ from the 
chromoelectric correlator $\GE$, one has to use the relation between this correlator
and the spectral function $\rho(\omega,T)$:
\begin{align}
\GE(\tau) &= \int_0^\infty \frac{\dd{\omega}}{\pi}\rho(\omega,T)K(\omega,\tauT)\,,\label{eq:gefromrho}\\
\intertext{where}\nonumber\\
K(\omega,\tauT) &= \frac{\cosh\left(\frac{\omega}{T}\left(\tauT-\frac{1}{2}\right)\right)}{\sinh\left(\frac{\omega}{2T}\right)}. \nonumber
\end{align}
The heavy quark diffusion coefficient is determined in terms of $\rho$ through the Kubo formula~\cite{Kapusta:2006pm}
\begin{equation}
\kappa \equiv \lim_{\omega\rightarrow 0} \frac{2T\rho(\omega,T)}{\omega}\label{eq:kappalimd}.
\end{equation}

At the leading order of the perturbation theory, the spectral function is given by~\cite{CaronHuot:2009uh}:
\be\label{eq:lorho}
\rho^\mathrm{LO}(\omega,T) = \frac{g^2(\mu_\omega)\CF\omega^3}{6\pi}\,,
\ee
where the coupling has been evaluated at the scale $\mu_\omega$. We use the five-loop running coupling constant
in this work~\cite{Tanabashi:2018oca}. 
At LO the scale $\mu_{\omega}$ is arbitrary.
A natural choice is $\mu_\omega^\mathrm{simple}=\mathrm{max}(\omega,\pi T)$~\cite{Francis:2015daa}.
The LO spectral function~\eqref{eq:lorho} gives $\kappa=0$.

At NLO, the perturbative calculation of $\rho(\omega,T)$ needs Hard-Thermal-Loop (HTL) resummation for
$\omega \lsim \mE$, with $\mE$ being the LO Debye mass: $\mE=\sqrt{\Nc/3} g T$ in the pure gauge theory. 
The full NLO result of $\rho(\omega,T)$ has been calculated in Ref.~\cite{Burnier:2010rp}.
The NLO spectral function provides the LO nonvanishing result for $\kappa$:
\be\label{eq:kappapert}
\frac{\kappa^\mathrm{LO}}{T^3} = \frac{g^4\CF \Nc}{18\pi}
\left[
\ln\frac{2T}{\mE}+\xi
\right]\,,
\ee
where $\xi = \frac{1}{2}-\gamma_\mathrm{E}+\frac{\zeta^\prime(2)}{\zeta(2)} \simeq -0.64718$.
For $\omega \gsim T$, there is no need for resummation when calculating the spectral function at NLO; 
the naive (nonresummed) NLO result for $\rho(\omega,T)$ in the pure gauge case reads \
\bea\label{eq:rho_nlo_t0}
 & &\rho_\mathrm{naive}(\omega,T)  = \\ & & 
 \frac{g^2 \CF \omega^3}{6\pi} \biggl\{ 1 + 
 \frac{g^2}{(4\pi)^2}\biggl[ \Nc \biggl( 
 \frac{11}{3} \ln\frac{\mu^2_\omega}{4\omega^2} 
 + \frac{149}{9} - \frac{8\pi^2}{3}
 \biggr) \biggr] \biggr\} 
 \nn & & +
 \frac{g^2 \CF}{6\pi} \frac{g^2}{2\pi^2} \biggl\{ 
 \Nc \int_0^\infty \! \dd{q} \, \nB{}(q) 
 \biggl[
   (q^2 +2 \omega^2) \ln \left| \frac{q+\omega}{q-\omega} \right| \nn & &
   + q \omega \biggl( \ln\frac{|q^2 - \omega^2|}{\omega^2} - 1 \biggr)
  \nn & & +
      \frac{\omega^4}{q} 
     \mathbbm{P} \biggl( \frac{1}{q+\omega} \ln \frac{q + \omega}{\omega} 
     +   \frac{1}{q-\omega} \ln \frac{\omega}{|q - \omega|} \biggr) \biggr]
 \biggr\}\,,\nonumber
\label{eq:rho_nlo_part}
\eea
where $\nB{}(q)=(\exp(q/T)-1)^{-1}$ is the Bose--Einstein distribution, 
$\mathbbm{P}$ takes the principal value,
and $g^2\equiv g^2(\mu_\omega)$. 
The first line of Eq.~\eqref{eq:rho_nlo_t0} gives the NLO $T=0$ contribution, and the subsequent lines
carry the thermal effects. 
For the NLO $\rho(\omega,T)$, $\mu_\omega$ may be set such that
the NLO $T=0$ contribution vanishes~\cite{Burnier:2010rp}:
\be\label{muom1}
\ln(\mu_\omega)=\ln(2\omega) +
\frac{(24\pi^2-149)}{66}\,,
\ee
and the $T=0$ part of Eq.~\eqref{eq:rho_nlo_t0} reduces to Eq.~\eqref{eq:lorho}.
This is a convenient choice of scale for $\omega \gg T$. 
For $\omega \sim T$ or smaller a convenient choice of scale was proposed in Ref.~\cite{Kajantie:1997tt}
\be\label{muom2}
\ln(\mu_\omega)= \ln(4\pi T) - \gamma_\mathrm{E} -\frac{1}{22}\,, 
\ee
in the pure gauge case. 
We switch between these two scales when they become equal at $\omega \simeq 0.8903\,T$~\cite{Burnier:2010rp}.

The heavy quark diffusion coefficient has been calculated at NLO, and the
result reads~\cite{CaronHuot:2008uh}:
\be\label{eq:kappaNLO}
\frac{\kappa^\mathrm{NLO}}{T^3} = \frac{g^4\CF \Nc}{18\pi}\left[
\ln\frac{2T}{\mE}+\xi+2.3302 \frac{\mE}{T}
\right].
\ee
The NLO result for $\kappa$ cannot be replicated from currently known spectral functions 
as that would require $\rho(\omega,T)$ to be available at NNLO, which it is not.
Both the LO and NLO results for $\kappa$ are obtained under the weak coupling assumption $\mE \ll T$. 
This condition, however, is not satisfied for most of the temperatures of interest.
As a consequence, one obtains an unphysical behavior at LO i.e., that $\kappa$ becomes negative for $T<10^3\,\Tc$. 

\begin{figure}[!ht]
  \includegraphics[width=8.6cm]{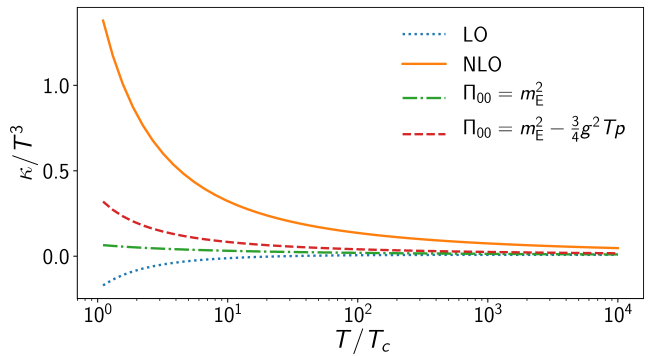}
  \caption[b]{Perturbative estimates of $\kappa$ for the pure gauge theory as a function of temperature 
   calculated at LO and NLO, as well as using the resummed leading-order expression~\eqref{eq:kappapertint}.
  }
  \label{fig:kappapertc}
\end{figure}

One can also calculate $\kappa$ using the kinetic theory. The corresponding expression
reads~\cite{CaronHuot:2007gq,CaronHuot:2008uh}:
\begin{align}\label{eq:kappapertint}
\kappa^\mathrm{LO} = &\frac{g^4 \CF}{12\pi^3}\int_0^\infty q^2\dd{q} \int_0^{2q} \frac{p^3\dd{p}}{(p^2 + \Pi_{00})^2} \\ 
&\times \Nc \nB{}(q)(1+\nB{}(q))
\left( 2-\frac{p^2}{q^2}+\frac{p^4}{4q^4}\right)\,. \nonumber
\end{align}
If we do not expand in the temporal gluon self-energy, $\Pi_{00}(p)$, 
which is formally of order $g^2$, the above expression contains higher-order contributions to $\kappa$ as well. 
Therefore, the above expression can be considered as the resummed leading-order result. 
The temporal gluon self-energy depends on the gauge choice.
For small momenta, it can be expanded as
\begin{equation}
\Pi_{00}(p)=\mE^2-\frac{\Nc}{4} g^2 T p+ ... \,.
\end{equation}
The first two terms in this expansion are gauge independent. We can take either the first term or the first
and second terms in the above expression and evaluate the integral in Eq.~\eqref{eq:kappapertint} numerically.
Only keeping the first term in the above expression for $\Pi_{00}$ already leads to a positive result, while
keeping the second term as well leads to an enhancement of the $\kappa$ value.
We present all the different perturbative results for $\kappa$ as a function of temperature in Fig.~\ref{fig:kappapertc}.
The scale of the coupling is the one defined in Eq.~\eqref{muom2}.
\begin{figure}[!ht]
  \includegraphics[width=8.6cm]{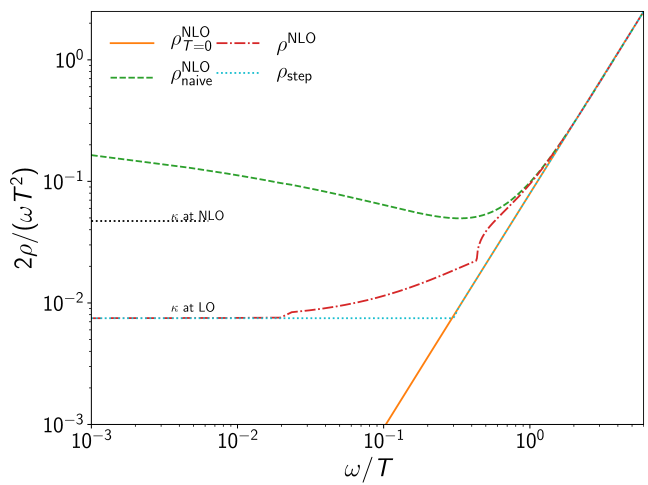}
  \caption[b]{The perturbative spectral functions $\rho(\omega,T)$ at $T=10^4\,\Tc$ for different orders of perturbation theory.
              The dotted lines on the left indicate the perturbative estimates of $\kappa$ given by Eqs.~\eqref{eq:kappapert} (LO)
              and~\eqref{eq:kappaNLO} (NLO).
  }
  \label{fig:spct104}
\end{figure}%
\begin{figure}[!ht]
  \includegraphics[width=8.6cm]{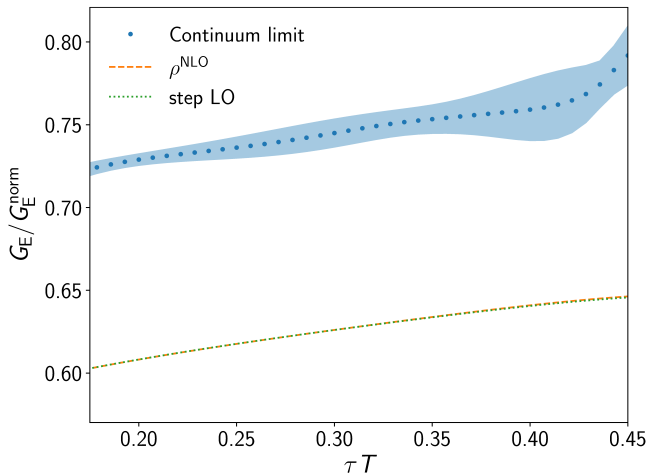}
  \caption[b]{The chromoelectric correlator at $T=10^4\,\Tc$ calculated from the NLO spectral function (orange band) 
              and the perturbative step form of the spectral function (green band). 
              The orange band completely overlaps with the green band, and it is hardly distinguishable from it.
              The errors for both orange and green bands come from varying the scale by a factor of 2.
              In blue, we show the continuum limit of the $T=10^4\,\Tc$ lattice data.
  }
  \label{fig:corrt104}
\end{figure}%

At the highest temperature considered in this, $T=10^4\,\Tc$, study we expect 
that the NLO result can provide some guidance on the properties of the spectral function 
and on the $\tau$ dependence of the chromoelectric correlator. 
Therefore, in Fig.~\ref{fig:spct104}  we show different versions of the NLO spectral function,
including the zero-temperature one. 
The full NLO spectral function can be described well by the simple $\kappa^{\mathrm{LO}} \omega/(2 T)$ form for $\omega<0.02\,T$, 
while it approximately agrees with the $T=0$ result for $\omega>2\,T$. 
The full NLO result and the naive (unresummed) NLO result agree for $\omega>0.6\,T$. 
At small $\omega$, the naive NLO result is logarithmically divergent. 
This divergence cancels against contributions coming from the scale $\mE$ in the resummed expression.
We can model the spectral function by smoothly matching the $\kappa^\mathrm{LO}\omega/(2T)$ behavior at small $\omega$ with
the zero-temperature spectral function at large $\omega$. 
We call this the perturbative step form.  
It is also shown in Fig.~\ref{fig:spct104} by the blue dotted line. 
By using the NLO spectral function evaluated for $\mu=\mu_{\omega}$, we can calculate the corresponding chromoelectric correlator, 
which is shown in Fig.~\ref{fig:corrt104}. Varying the renormalizarion scale by a factor of 2 around $\mu_{\omega}$ leads only to
very small changes of the correlator, roughly corresponding to the width of the line in Fig.~\ref{fig:corrt104}.
We also calculate the chromoelectric correlator corresponding to the perturbative step form.
The resulting correlator is indistinguishable from the one obtained using the NLO spectral function.
This means that the additional structures in the spectral function in the region $0.02 < \omega/T < 0.6 $ play no significant role
when it comes to the correlator.
We have also considered a perturbative step model using $\kappa^{\mathrm{NLO}}$. 
While using the NLO result for $\kappa$ significantly
enhances the spectral function in the low $\omega$ region it only leads to a $0.2\%$ enhancement of the 
chromoelectric correlator compared to the one obtained using $\kappa^{\mathrm{LO}}$. Thus, the correlator is not
sensitive to the small $\omega$ part of the spectral function at the highest temperature. 
At lower temperatures, $\kappa/T^3$ gets larger, and the contribution of the low $\omega$ part of the spectral functions is 
more prominent. 
Therefore, it is at lower temperatures that  the value of $\kappa$ can be constrained 
by accurate calculations of the chromoelectric correlator.

While at  $T=10^4\,\Tc$, one may expect the resummed NLO result to provide an adequate description of the spectral function, 
this is not expected at lower temperatures, because, as pointed out above, numerically $\mE>T$. In particular,
for $T<10^3\,\Tc$ the resummed spectral function turns negative at some point in the region $\omega<T$, 
thus implying that the resummed perturbative  result
is not applicable in this $\omega$ range. In Sec.~\ref{sec:model} we will discuss the implications of this finding.

In Fig.~\ref{fig:corrt104}, we also show the continuum limit of the chromoelectric correlator
at the high temperature $T=10^4\,\Tc$ for comparison. 
The continuum-extrapolated lattice result of the chromoelectric correlator
has the same shape as the NLO calculation.
We note, however, that our continuum data differ from the perturbative curve 
by a factor 1.2, which indicates that the renormalization constant is not accurate.
If we normalize the above lattice result to the correlator obtained from the NLO spectral function discussed
above at $\tau T=0.19$, we find that the two agree within errors.

\section{Short-time behavior of the lattice results on the electric correlator}\label{sec:cont}

The continuum results of $\GE$ normalized by $\GEnorm$ show significant dependence on $\tau$. 
The analysis in Sec.~\ref{sec:meas} implies that this cannot be caused by thermal effects 
(cf. Figs.~\ref{fig:fermipert} and~\ref{fig:therratio} ).  The LO result does not take into
account the effect of the running of the gauge coupling, and this could be the reason
why $\GE^\mathrm{LO}$, or equivalently $\GEnorm$ (which is the same up to a multiplicative factor)
does not capture the $\tau$ dependence of the chromoelectric correlator.
Therefore, as an alternative normalization we consider a correlator obtained from Eq.~\eqref{eq:gefromrho}
using the zero-temperature NLO result for the spectral function with a running coupling constant evaluated
at scale $\mu_{\omega}$ given by Eqs.~\eqref{muom1} and~\eqref{muom2}.
We label the corresponding correlator
as $\GE^\mathrm{NLO+}$. 
\begin{figure}[!ht]
  \includegraphics[width=8.6cm]{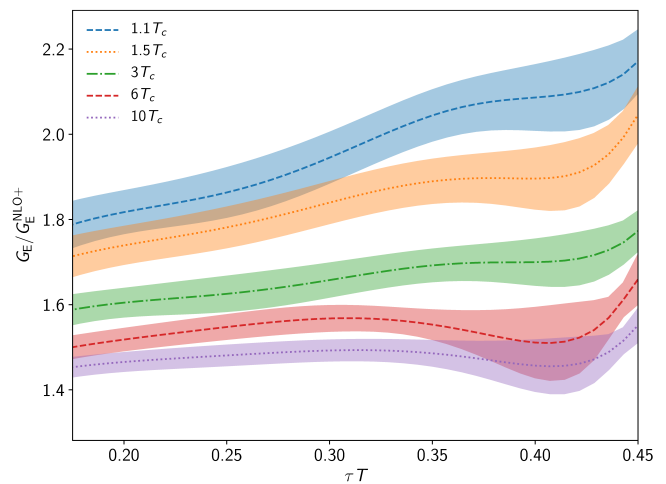}
  \caption[b]{Continuum extrapolation for all temperatures as a function of $\tauT$.
  }
  \label{fig:tempg2norm}
\end{figure}%
\begin{table}
  \begin{ruledtabular}
   \begin{tabular}{c|cccccc}
   \diagbox{$\mu$}{$\Tc$}  & 1.1  & 1.5  & 3        & 6        & 10       & $10^4$   \\ \hline
   0.5$\mu_\omega$ & 1.82(5)  & 1.74(5)  & 1.61(3)  & 1.52(3)  & 1.47(2)  & 1.20(1)  \\
   1$\mu_\omega$   & 1.81(5)  & 1.73(5)  & 1.60(3)  & 1.51(3)  & 1.46(2)  & 1.20(1)  \\
   2$\mu_\omega$   & 1.84(5)  & 1.76(5)  & 1.62(3)  & 1.53(3)  & 1.48(2)  & 1.20(1)  \\
   \end{tabular}
  \end{ruledtabular}
  \caption{Normalization factor $C_\mathrm{N}$ for three different renormalization scales (rows)
   and for each measured temperature (columns) at $\tauT=0.19$.
   }
  \label{tab:cn}
\end{table}%
The numerical results for $\GE/\GE^\mathrm{NLO+}$ are shown in Fig.~\ref{fig:tempg2norm}. 
We see that this ratio increases less with increasing $\tau$.
We expect that for small enough $\tau T$, one would see a plateau in $\GE/\GE^\mathrm{NLO+}$.
This is not the case, however, because the continuum extrapolation is not reliable for $\tau T<0.17$, and
for larger $\tau T$ there are some small thermal effects. On the other hand, given the uncertainties, 
we could still fit  $\GE/\GE^\mathrm{NLO+}$ with a constant for $0.17 \le \tau T \le 0.19$.
This indicates that $\GE^\mathrm{NLO+}$ captures the $\tau$ dependence of the chromoelectric correlator 
obtained on the lattice much better. 
However, even at the smallest $\tau$ the ratio $\GE/\GE^\mathrm{NLO+}$ is different from 1. 
This is most likely due to the fact that the one-loop result is not accurate for $\ZE$. 
As shown in the previous section, even for the highest temperature, 
$T=10^4\,\Tc$, the NLO result is lower by a factor of 1.2, as seen in Fig.~\ref{fig:corrt104}, 
although its $\tau$-dependence agrees well with the continuum extrapolated lattice data. 
Therefore, we introduce an additional normalization factor, 
$C_\mathrm{N}$ by normalizing the ratio $\GE/\GE^\mathrm{NLO+}$ to 1 at $\tauT=0.19$. 
To check the uncertainty of $C_\mathrm{N}$ due to the choice of the normalization point, 
we also consider $\tauT=0.175$ as a possible normalization point. 
Furthermore, we vary the scale $\mu_{\omega}$ by a factor of 2 around the optimal value when evaluating $C_\mathrm{N}$.
The numerical values of $C_\mathrm{N}$ are shown in Tab.~\ref{tab:cn} for different temperatures. 
The dependence on the normalization point is shown in the systematic error and is of the same order as the scale dependence. 
The additional normalization constant $C_\mathrm{N}$ decreases with increasing temperature. 
This is due to the fact that the $\beta$ range used in the evaluation of the lattice correlator is increasing 
with increasing temperature, and the one-loop result is more reliable at large $\beta$ values.
We will normalize $\GE/\GE^\mathrm{NLO+}$ with $C_\mathrm{N}$ given in Table~\ref{tab:cn} 
before comparing with the model spectral functions used for the extraction of $\kappa$.  

\section{Modeling the spectral function and determination of $\kappa$}\label{sec:model}

To obtain the heavy quark diffusion coefficient from the continuum-extrapolated lattice results, we need to assume
some model for the spectral function. We will use the NLO results on the spectral function as well as $\kappa$
to guide us in this process. We also need to consider how sensitive the Euclidean-time chromoelectric
correlator is to the spectral function in different $\omega$ regions. From the previous sections it is clear
that $G_E$ is dominated by the large $\omega$ part of the spectral function and thermal effects in the 
spectral function contribute at the level of a few percent to the correlator.

It is reasonable to assume that at large enough $\omega$, perturbation theory is reliable even if
the condition $m_E \ll T$ is not satisfied. This is because for large $\omega$ HTL resummation is not
important, as will be detailed later. Certainly at zero temperature the perturbative calculation of $\rho(\omega,T)$ is
reliable for $\omega \gg \Lambda_\mathrm{QCD}$. Therefore, we assume that for $\omega>\omega^\mathrm{UV}$ the spectral function
is given by $\rho^\mathrm{UV}(\omega,T)$, which is calculated perturbatively. On the other hand, for sufficiently small
$\omega$, the spectral function is given by
\be
\rho^\mathrm{IR}(\omega,T) = \frac{\omega\kappa}{2T},\label{eq:rhoIR}
\ee
and we can assume that $\rho(\omega,T)=\rho^\mathrm{IR}(\omega,T)$ for $\omega<\omega^\mathrm{IR}$.
In the region $\omega^\mathrm{IR}<\omega<\omega^\mathrm{UV}$, the form of the spectral function is not known, in general,
and this lack of knowledge will generate an uncertainty in the determination of $\kappa$. 
Based on these considerations, we adopt the following procedure to estimate $\kappa$:
\begin{itemize}
\item
For a given value of $\kappa$, we construct the model spectral function that is given by the NLO result at high energy, $\omega > \omega^\mathrm{UV}$, and
by Eq. \eqref{eq:rhoIR} at low energy, $\omega< \omega^\mathrm{IR}$. 
\item
For intermediate $\omega$, namely $\omega^\mathrm{IR} \le \omega \le \omega^\mathrm{UV}$, we consider various forms of the spectral function 
such that the total spectral function is smooth, as described below.
\item
We match the continuum-extrapolated lattice result for the chromoelectric correlator to the correlator obtained from the model spectral
function at small $\tauT$ and adjust the value of $\kappa$ to obtain the best description of the lattice result.
We take care that $\tauT$ is not too small, so that the procedure is not affected by lattice artifacts.
\item 
  We estimate the uncertainties due to modeling the spectral function for $\omega^\mathrm{IR} \le \omega \le \omega^\mathrm{UV}$,
  due to the normalization point in $\tauT$, and
  due to the choice of the renormalization scale in the NLO result.
\end{itemize}

We consider 
two possible forms of the spectral functions that are continuous and are 
based on simple interpolations between the small $\omega$ (IR)
region and the large $\omega$ (UV) region:
\begin{align}\label{eq:ans-lin}
&\rho_\mathrm{line}(\omega,T) = \rho^\mathrm{IR}(\omega,T)\theta(\omega^\mathrm{IR}-\omega) +\\ 
&\left[\frac{\rho^\mathrm{IR}(\omega,T)-\rho^\mathrm{UV}(\omega,T)}{\omega^\mathrm{IR}-\omega^\mathrm{UV}}
\left(\omega-\omega^\mathrm{IR}\right)+\rho^\mathrm{IR}(\omega,T)\right] \times \nonumber \\ 
&\theta(\omega - \omega^\mathrm{IR})\theta(\omega^\mathrm{UV}-\omega) + \rho^\mathrm{UV}(\omega,T)\theta(\omega-\omega^\mathrm{UV})\,, \nonumber
\end{align}
and
\be\label{eq:ans-step}
\rho_\mathrm{step}(\omega,T) = \rho^\mathrm{IR}(\omega,T) \,\theta(\Lambda-\omega) + 
\rho_{T=0}^\mathrm{UV}(\omega,T)\,\theta(\omega-\Lambda)\,.
\ee
The latter case corresponds to $\omega^\mathrm{IR}=\omega^\mathrm{UV}=\Lambda$ 
and the value of $\Lambda$ is self-consistently determined
by the continuity of the spectral function for a given $\kappa$. Thus, this model depends only on $\kappa$.
In the former case additional considerations are needed to fix $\omega^\mathrm{IR}$ and $\omega^\mathrm{UV}$, 
which are described below.
We will refer to these two forms as the line model and the step model, respectively.

The NLO result for the spectral function naturally interpolates
between the IR and UV regions, but it is not reliable for small $\omega$ even at the highest temperature
as discussed in Sec.~\ref{sec:rho}. However, it
can provide some guidance on how to choose $\omega^\mathrm{IR}$ and $\omega^\mathrm{UV}$. As mentioned above,
for $\omega>T$, HTL resummation may not be important, and the naive and resummed NLO result for
the spectral function should agree. As discussed in  Appendix~\ref{sec:appB}, the resummed and naive NLO results
for the spectral function agree well for $\omega>2.2\,T$. Furthermore, the thermal contribution
to $\rho(\omega,T)$ is about the same for $\omega>2.2\,T$ at the lowest and the highest temperature 
when normalized by $\omega T^2$. This indicates that the perturbative calculations are reliable
for these values of $\omega$. Therefore, we choose $\omega^\mathrm{UV}=2.2\,T$. 
At the highest temperatures, the resummed NLO result is well described by the linear form given by Eq.~\eqref{eq:rhoIR}
with $\kappa=\kappa^\mathrm{LO}$ for $\omega<0.02\,T$. Therefore, $\omega^\mathrm{IR}=0.01\,T$ appears to be
a reasonable choice. 
The NLO result for $\kappa$ is significantly larger than the LO result, implying that the spectral
function at low $\omega$ is also larger and therefore will match $\rho^\mathrm{UV}(\omega,T)$ at larger
$\omega$. We find that $\rho^\mathrm{IR}(\omega,T)$ and $\rho^\mathrm{UV}(\omega,T)$  are equal 
at around $\omega=0.4\,T$. Therefore, besides $\omega^\mathrm{IR}=0.01\,T$, we will also use 
$\omega^\mathrm{IR}=0.4\,T$ and $\omega^\mathrm{IR}=1\,T$ in our analysis.

\begin{figure}[!ht]
  \includegraphics[width=8.6cm]{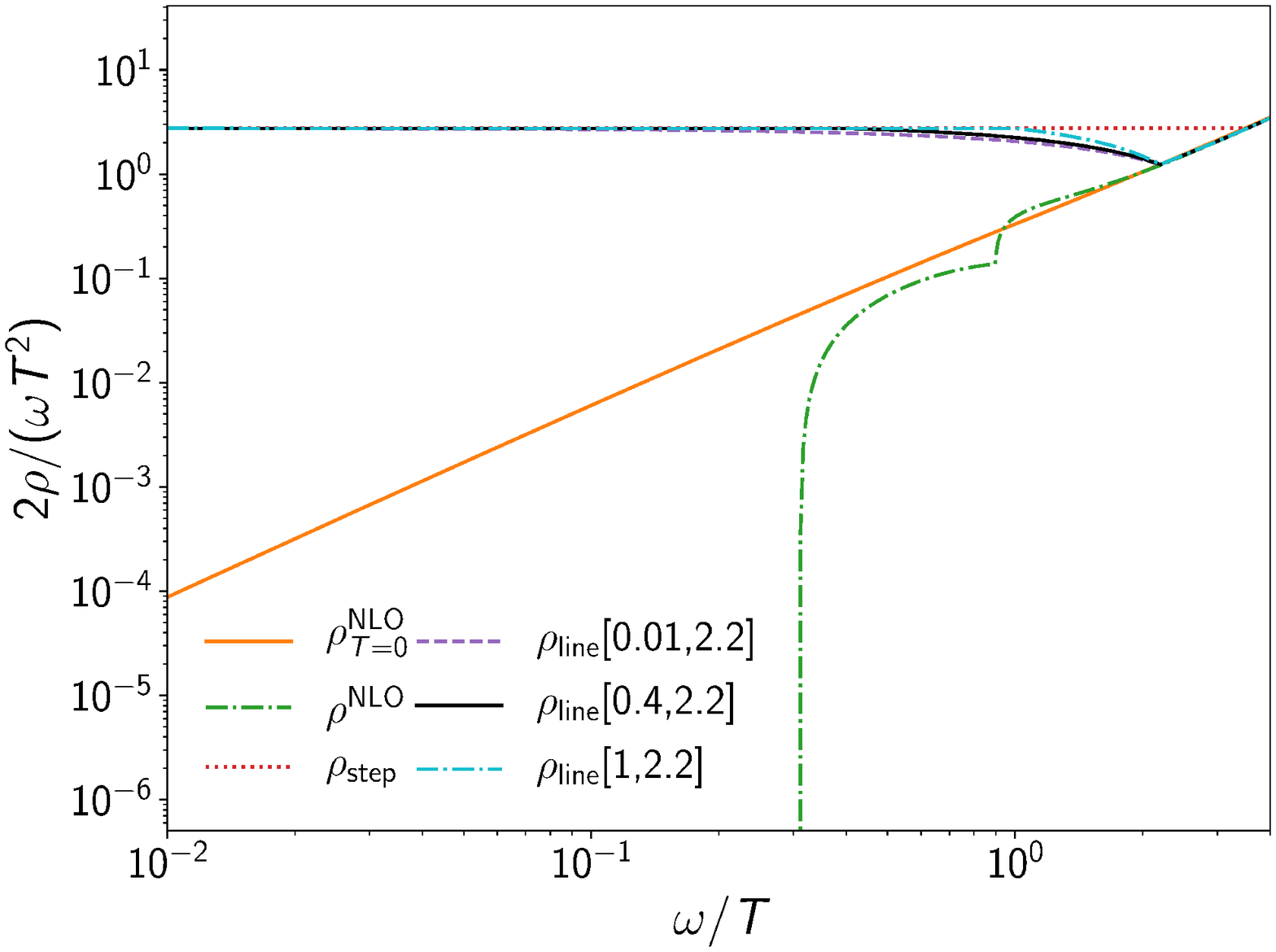}
  \includegraphics[width=8.6cm]{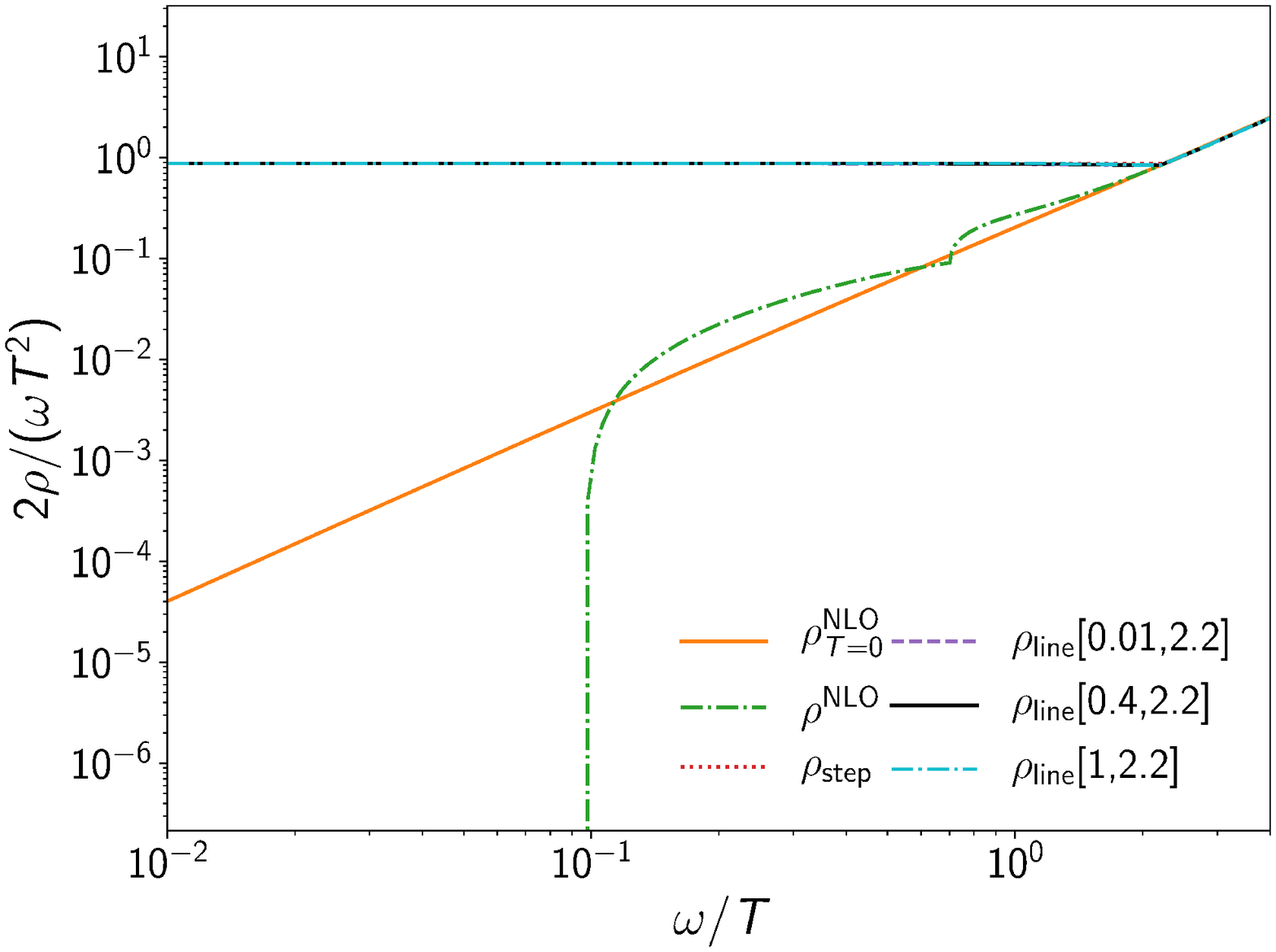}
  \includegraphics[width=8.6cm]{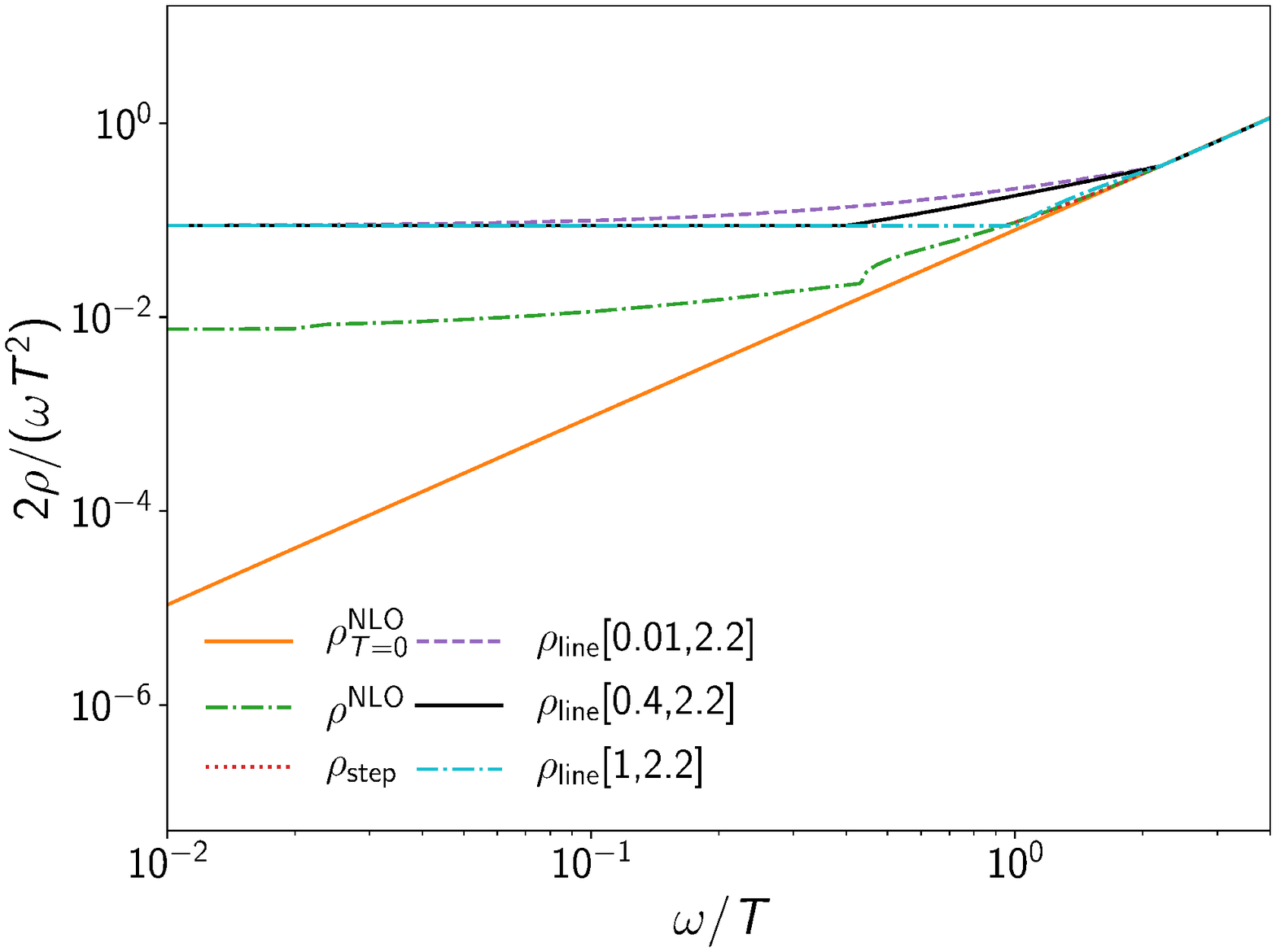}
  \caption[b]{The shapes of different spectral function models $\rho(\omega,T)$ at (from top to bottom) $T=1.1\,\Tc$, $T=6\,\Tc$, and $T=10^4\,\Tc$.
The arguments of $\rho_\mathrm{line}$ in square brackets stand for 
$[\omega^\mathrm{IR},\omega^\mathrm{UV}]$.
  }
  \label{fig:spf_mod} 
\end{figure}

In Fig.~\ref{fig:spf_mod} we show the spectral functions obtained from Eqs.~\eqref{eq:ans-lin}
and~\eqref{eq:ans-step}, assuming $\kappa=\kappa^\mathrm{NLO}$ in $\rho_\mathrm{step}$ and $\rho_\mathrm{line}$,
and three different $\omega^\mathrm{IR}$
at three representative temperatures, $T=1.1\,\Tc,~6\,\Tc$, and $10^4\,\Tc$. 
From the figure, we see that at the lowest temperature, the $\rho_\mathrm{step}(\omega,T)$ model matches
the UV behavior at larger $\omega$ without the dip around $\omega\sim T$ of
the $\rho_\mathrm{line}(\omega,T)$ model.
The $\rho_\mathrm{line}$ form with $\omega^\mathrm{IR}=0.01\,T$ and $\rho_\mathrm{step}$ provide
upper and lower bounds for the spectral function at $T=1.1\,\Tc$. The picture is the same 
for $T=1.5\,\Tc$ and $T=3\,\Tc$. At $T=6\,\Tc$, all forms of the spectral functions provide nearly
identical results. At the highest two temperatures, the possible choices of the spectral functions
are limited by $\rho_\mathrm{line}$ with $\omega^\mathrm{IR}=0.01\,T$ and $\omega^\mathrm{IR}=T$. 

Using the models for the spectral functions described above, we have calculated the corresponding
Euclidean-time chromoelectric correlators for different values of $\kappa$ and compared these
with the continuum-extrapolated lattice results at each temperature to estimate the heavy
quark diffusion coefficient. As discussed in the previous section, the continuum-extrapolated lattice
results need an additional renormalization because the one-loop renormalization constant, $Z_E$, is not 
accurate. Therefore, we have matched the correlator obtained from the model spectral function 
to the continuum-extrapolated lattice data at $\tauT=0.19$. 
The resulting multiplicative constants $C_N$ are slightly different from those shown in Table~\ref{tab:cn}.
This is because the correlators obtained from the model spectral functions are slightly different
from $\GE^\mathrm{NLO+}$ at  $\tauT=0.19$ due to the thermal contribution.
We demonstrate this procedure in Appendix~\ref{sec:appB} for different model spectral functions.
Different forms give different values of $\kappa$, and this is the dominant source of systematic
error in the determination of $\kappa$.
We have also studied the dependence of $\kappa$ on the choice of the normalization point in $\tau$ and
the choice of the renormalization scale. Choosing the normalization point in the range $0.17 \le \tauT  \le 0.19$
leads to an 8\% variation in the resulting $\kappa$. Varying the renormalization scale by a factor of 2 results
in a similar variation. 

Putting everything together we obtain the following estimates for the heavy quark diffusion coefficient from the
analysis:
\bea
1.91 < \frac{\kappa}{T^3} < 5.4  \;&\text{for}&\; T=1.1\,\Tc\,, \label{eq:kappa11} \\
1.31 < \frac{\kappa}{T^3} < 3.64  \;&\text{for}&\; T=1.5\,\Tc\,, \label{eq:kappa15}\\
0.63 < \frac{\kappa}{T^3} < 2.20  \;&\text{for}&\; T=3\,\Tc\,,\label{eq:kappa3} \\
0.43 < \frac{\kappa}{T^3} < 1.05      \;&\text{for}&\; T=6\,\Tc\,,\label{eq:kappa6} \\
0 < \frac{\kappa}{T^3}    < 0.72     \;&\text{for}&\; T=10\,\Tc\,,\label{eq:kappa10} \\
0 < \frac{\kappa}{T^3}    < 0.10    \;&\text{for}&\; T=10^4\,\Tc\,,\label{eq:kappa14}
\eea
although one should be reminded that, as discussed at the end of Sec.~\ref{sec:rho}, 
the lattice data are weakly sensitive to $\kappa$ at the highest temperature.
The dominant uncertainty in the above result comes from the form of the spectral function used in the analysis and
the uncertainty of the continuum-extrapolated lattice results.

\begin{figure}[!ht]
  \includegraphics[width=8.6cm]{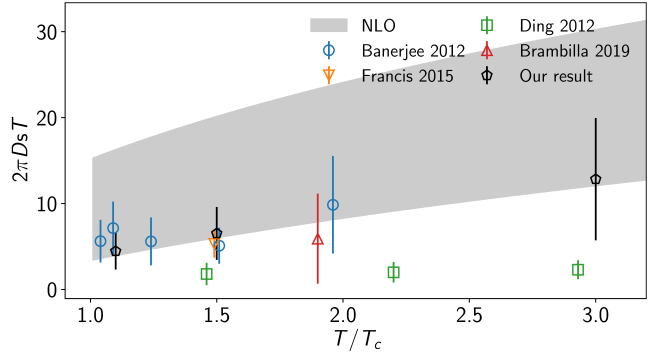}
  \caption[b]{Our results compared to existing lattice studies.
              The shaded band shows the perturbative behavior~\eqref{eq:kappaNLO} 
              and the effect of the scale $\mu_\omega$ being varied by a factor of 2.
  }
  \label{fig:resultothers} 
\end{figure}

We compare our result on $\kappa$ with the results of other lattice studies 
\cite{Ding:2012sp,Meyer:2010tt,Francis:2011gc,Banerjee:2011ra,Francis:2015daa}
in terms of 
the spatial diffusion coefficient $D_s$, which is given by the relation $\kappa/T^3 = 2/(D_s T)$,
in the temperature range $\Tc-3\,\Tc$. 
This is shown in Fig.~\ref{fig:resultothers}.
We see that our results agree well with the other lattice determinations, with the exception
of the one in Ref.~\cite{Ding:2012sp} that is based on charmonium correlators. This is likely due
to the fact that the determination of $D_s$ from the quarkonium correlators is not accurate, since the width
of the transport peak is difficult to determine \cite{Petreczky:2005nh,Petreczky:2008px}.

\begin{figure}[!ht]
  \includegraphics[width=8.6cm]{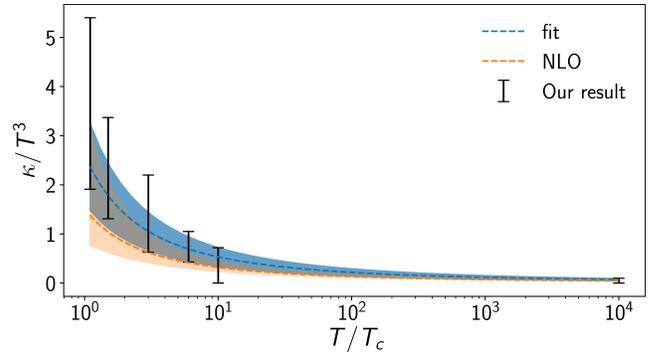}
  \caption[b]{Temperature dependence of our results compared to the NLO result. 
              The shaded bands include the errors coming from varying the scale by a factor of 2.
              The blue band also includes the statistical error.
  }
  \label{fig:fitkappac} 
\end{figure}

The temperature dependence of the heavy quark diffusion coefficient in the entire temperature region
is shown in Fig.~\ref{fig:fitkappac}. We clearly see the temperature dependence of $\kappa^3/T$. 
The $\kappa$ obtained on the lattice is not incompatible with the NLO result given the large errors. 
Inspired by this, we fit the temperature dependence of the lattice result
by modeling 
it on Eq.~\eqref{eq:kappaNLO} but keeping the coefficient of $\mE/T$ as a free parameter $C$.
From the fit, we obtain $C=3.81(1.33)$, which is larger than the NLO perturbative result $C\approx 2.3302$.

We note that our result is significantly larger than the simple holographic estimate~\cite{Kovtun:2003wp}: $2\pi D_s T= 1$.
However, more recent holographic estimates~\cite{Andreev:2017bvr} are close to our results.
Finally, comparing with more experimental quantities, we note that our result for $D_s$ at the lowest temperature 
is in agreement with the calculations of $D$~\cite{Tolos:2013kva} 
and $\bar{\mathrm{B}}$~\cite{Torres-Rincon:2014ffa} mesons propagating in a medium of light hadrons,
which find $2 \pi D_s T \sim 5$ for $T \approx \Tc$,
but much smaller than an earlier pion gas study~\cite{Abreu:2011ic} that found $2 \pi D_s T \approx 17$ for $T \approx \Tc$.
Experimental determinations of the D-meson azimuthal anisotropy coefficient $\nu_2$ 
at ALICE~\cite{Acharya:2017qps} and STAR~\cite{Adamczyk:2017xur} estimate at $T\approx \Tc$ that
$\kappa/T^3 \approx 1.8-8.38$ and $\kappa/T^3 \approx 1.05 - 6.28$, respectively.
These are in agreement with our findings.
All these experimental determinations include mass-dependent contributions,
while our determination of $\kappa$ is in the heavy quark limit.
Therefore the two should agree up to $1/m$ corrections. 

\section{Conclusions}\label{sec:concl}

In this paper, we have studied the chromoelectric correlator, 
$\GE$ at finite temperature on the lattice with the aim of extracting the heavy quark diffusion coefficient $\kappa$. 
The calculations have been performed in quenched QCD (SU(3) gauge theory) in order 
to obtain small statistical errors with the help of the multilevel algorithm. 
We have studied the dependence of the chromoelectric correlator on the Euclidean time, $\tau$, in
a wide temperature range in order to enable the comparison with weak coupling results.
It turned out that the $\tau$-dependence of the electric correlator is poorly captured by
the leading order result. Going beyond the leading-order result and incorporating the
effect of the running coupling in the corresponding spectral function results in a correlation
function $\GE^\mathrm{NLO+}$ that can capture the $\tau$-dependence of the lattice result much better.

To fully describe the $\tau$-dependence of $\GE$ calculated on the lattice, 
the effect of $\kappa$ encoded in the low $\omega$ part of the chromoelectric spectral function has to be considered. 
At high $\omega$, we have used forms of the spectral function that are motivated
by the next-to-leading-order perturbative results. 
Fitting 
the lattice results on $\GE$, we have obtained values of $\kappa$ at
different temperatures. We observe that the sensitivity of the chromoelectric correlator 
to $\kappa$ is small, varying from a few percent at the lowest temperatures to sub-percent
at the highest temperatures. This finding is corroborated by a model-independent analysis
of the chromoelectric correlator, cf. Figs.~\ref{fig:fermipert} and~\ref{fig:therratio}.
It is this small sensitivity that makes the lattice determination of $\kappa$ quite challenging.
Our main result is summarized
in Fig.~\ref{fig:fitkappac}, which shows the temperature dependence of the heavy quark diffusion
coefficient. For $T<2\,\Tc$, our results agree with other lattice determinations, while at higher
temperatures, they appear consistent with the NLO result.

One of the shortcomings of the present analysis is the use of the one-loop result for the renormalization
constant, $\ZE$, which, as we argued, is not reliable. The use of tadpole-improved one-loop perturbation
theory will not help, since it will make $\ZE$ even larger, while in order to achieve agreement of the lattice
and NLO results at small $\tau$, we need a smaller $\ZE$ than the one-loop result. Clearly, a nonperturbative
renormalization procedure will be needed, but this is beyond the scope of the present paper.

\acknowledgments{
We thank Saumen Datta for providing the code for the lattice calculation
of the chromo-electric correlator.
V.L. thanks Mikko Laine for clarifying details on the numerical evaluation of the NLO spectral function.
N.B., V.L., and A.V. acknowledge the support from 
the Bundesministerium f\"ur Bildung und Forschung project no.~05P2018 and by 
the DFG cluster of excellence \href{www.origins-cluster.de}{ORIGINS}
funded by the 
Deutsche Forschungsgemeinschaft  
under Germany's Excellence Strategy - EXC-2094-390783311.
P.P. has been supported 
by the U.S. Department of Energy under Contract No.~DE-SC0012704.
The simulations have been carried out 
on the computing facilities of the Computational 
Center for Particle and Astrophysics (C2PAP) of the cluster of excellence ORIGINS.
}

\appendix
\section{Infinite-volume limit and continuum extrapolation}\label{sec:appA}

\begin{figure*}
  \includegraphics[width=8cm]{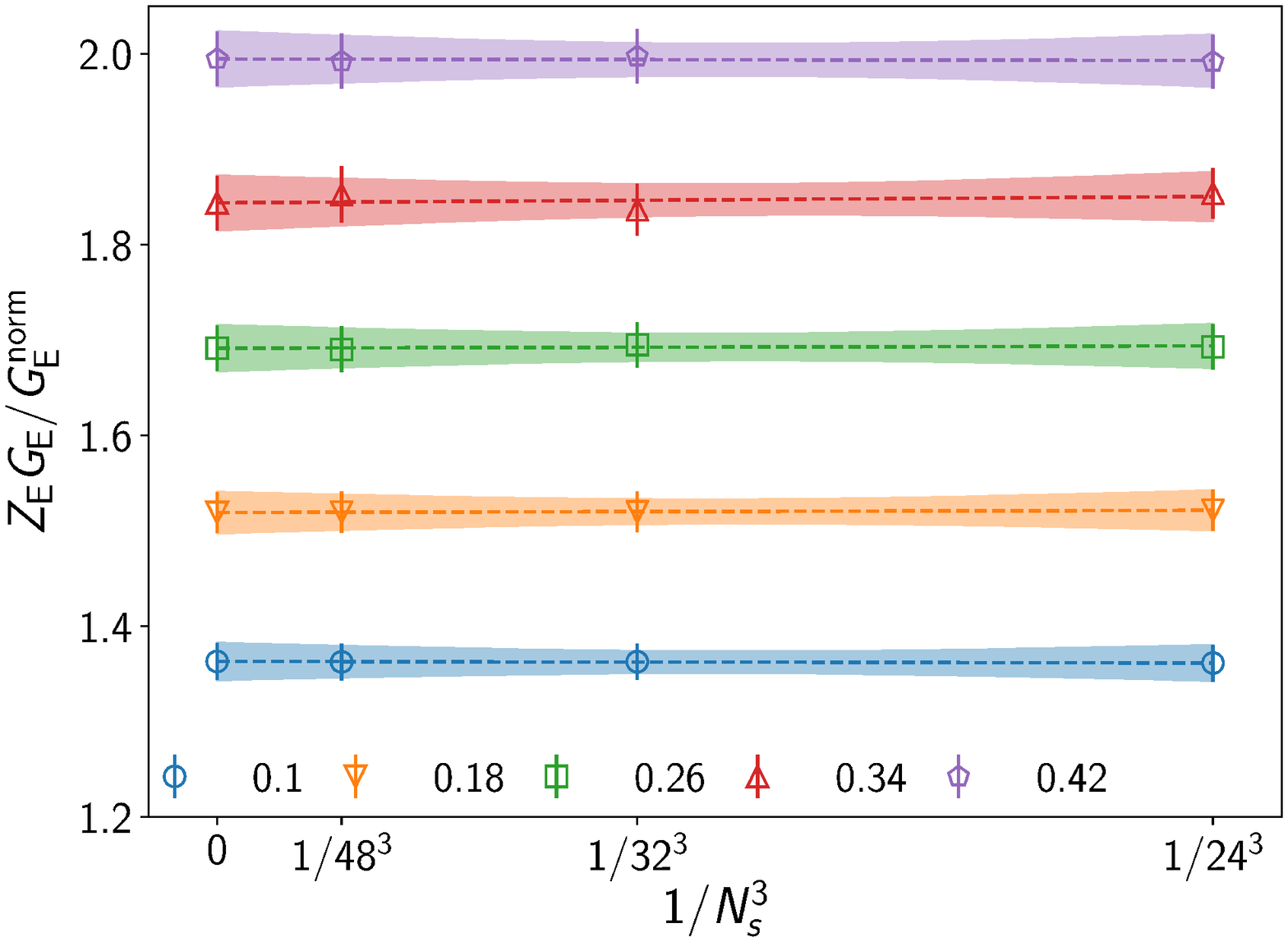}
  \includegraphics[width=8cm]{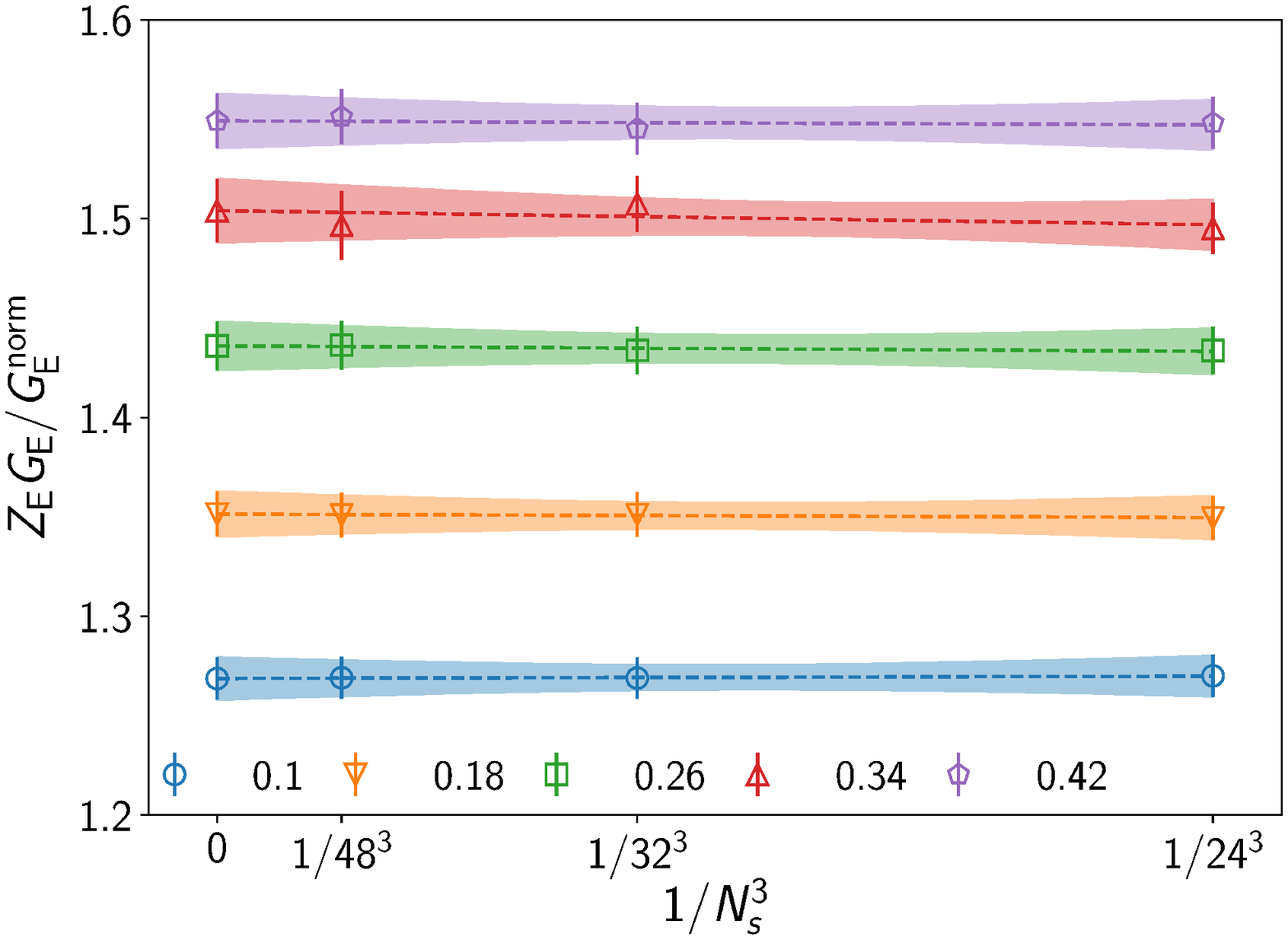} 
  \caption[b]{ Finite-volume effects for several $\tauT$ values presented with different colors 
               at $T=1.5\,\Tc$ (left) and $T=10\,\Tc$ (right). 
               The lines and bands correspond to the $1/\Ns^2$ fits and their uncertainties.
  }
  \label{fig:voleff} 
\end{figure*}

To check to what extent using lattices with an aspect ratio $\Ns/\Nt$ smaller than 4 leads to visible 
finite-volume effects, we have performed calculations at two temperatures, $T=1.5\,\Tc$ and $T=10\,\Tc$, on $\Ns^3 \times 12$
lattices with $\Ns=24,~32$, and $48$. 
The numerical results are shown in Fig.~\ref{fig:voleff} for some representative values of $\tauT$.
As one can see from the figure, the finite-volume effects are small. 
We have also attempted to perform an infinite-volume extrapolation by fitting the lattice results with a
$1/N_s^3$ form. The corresponding fits are shown in the figure as lines and bands together
with the infinite-volume result. It is clear from the figure that the differences between 
the infinite-volume result and the lattice results with different $\Ns$ values are of the order of the statistical errors.
Therefore, the use of $\Ns=48$ is justified.

\begin{figure*}
\includegraphics[width=8cm]{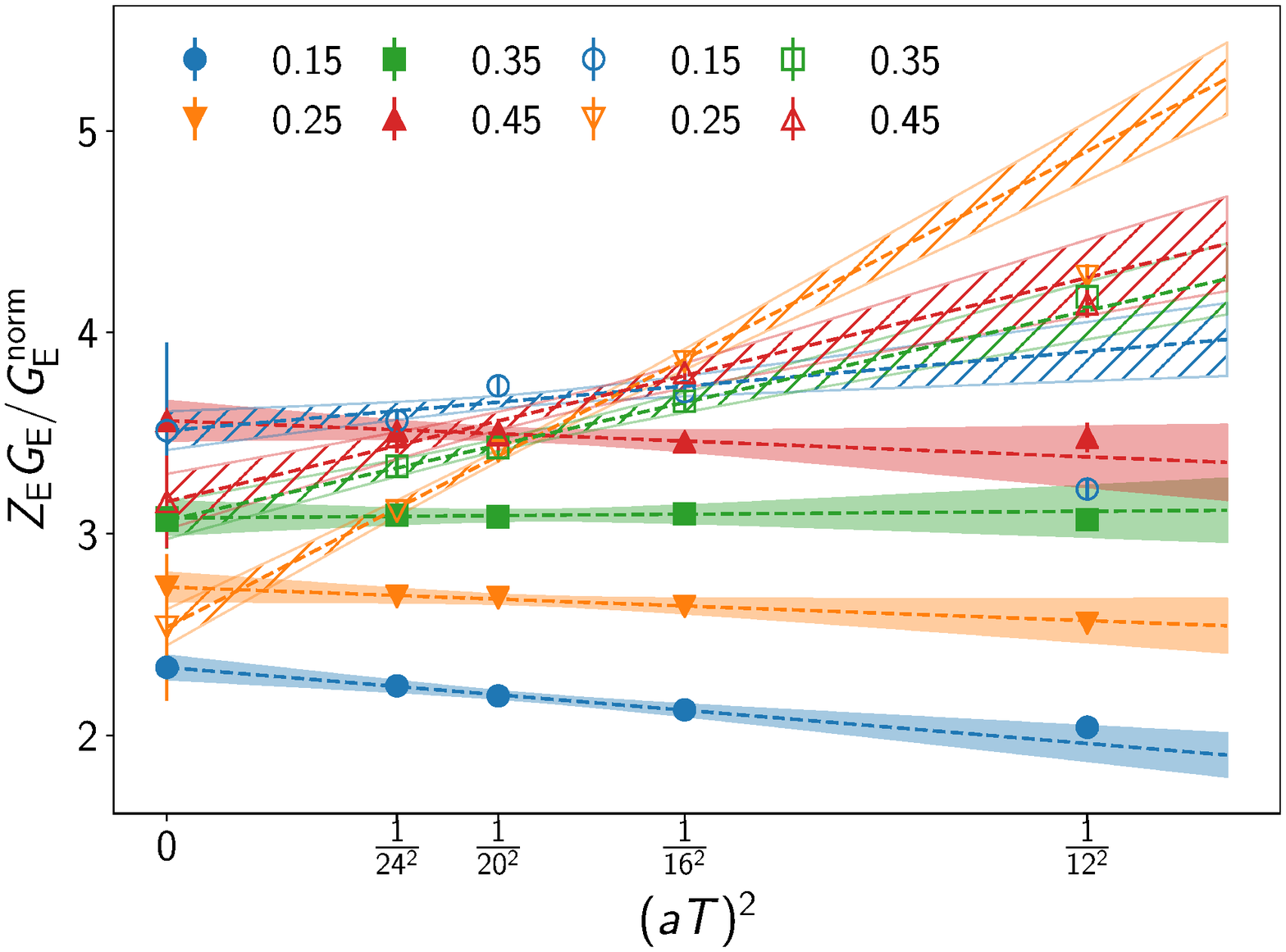}
\includegraphics[width=8cm]{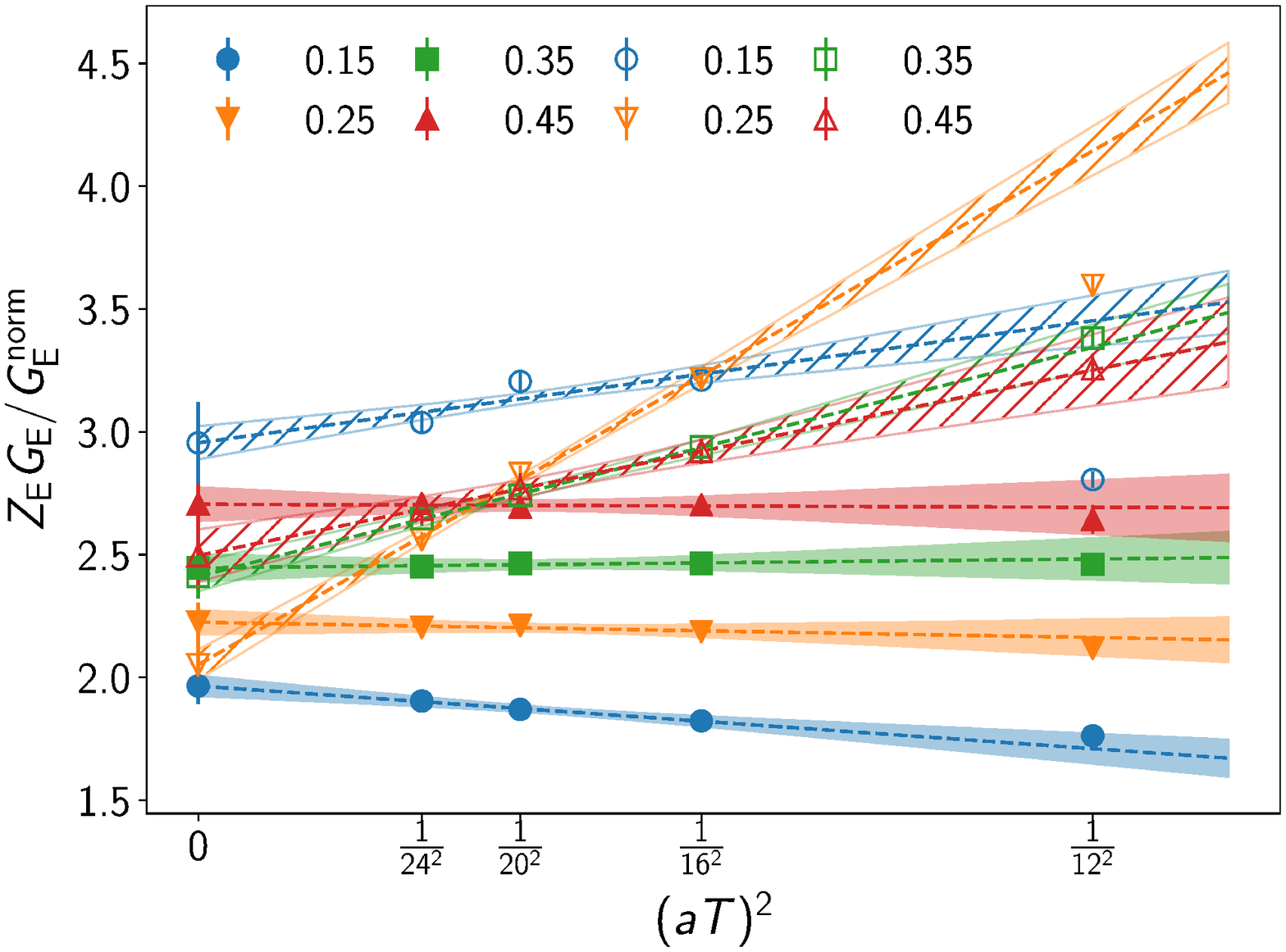}
\includegraphics[width=8cm]{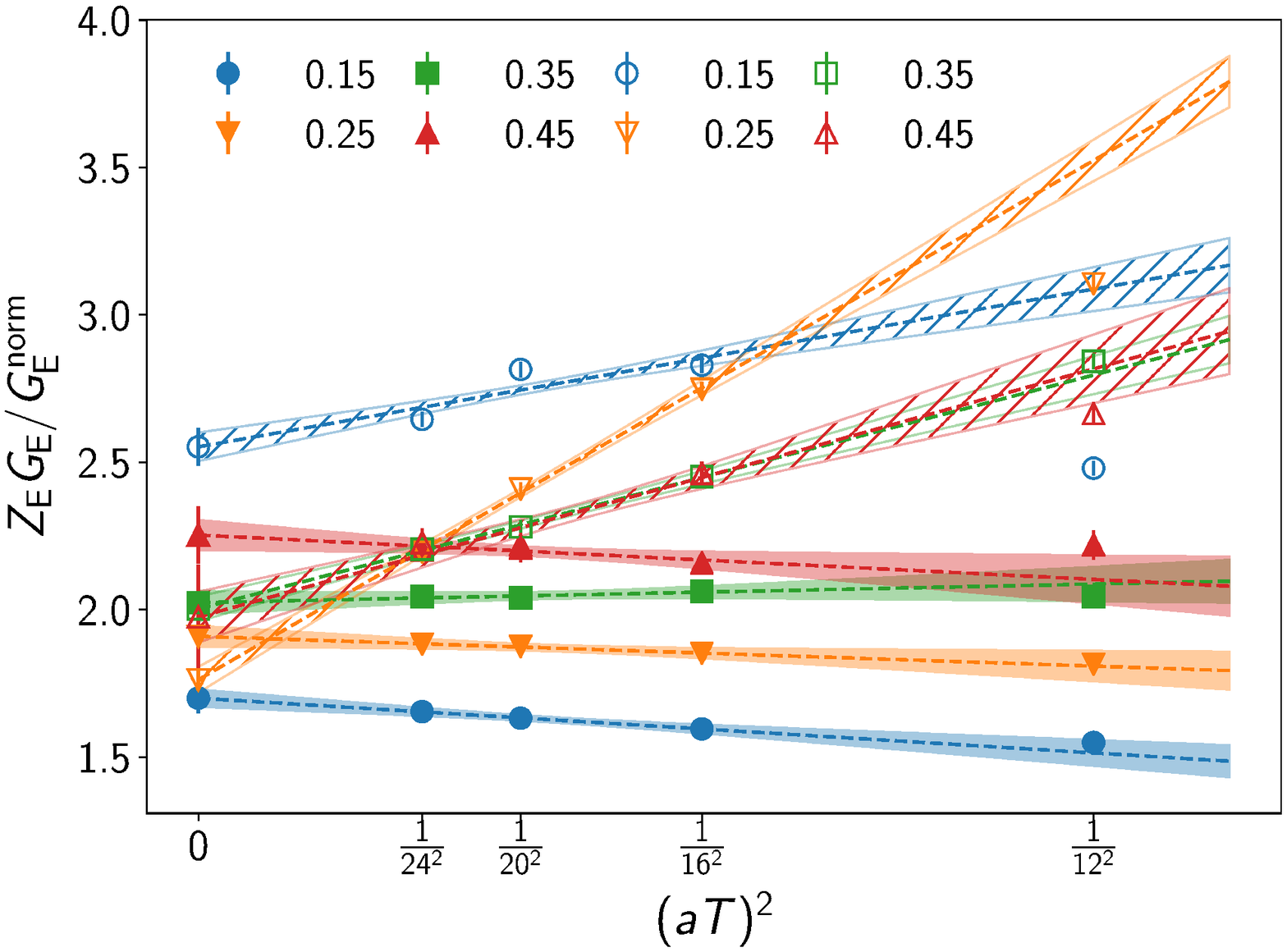}
\includegraphics[width=8cm]{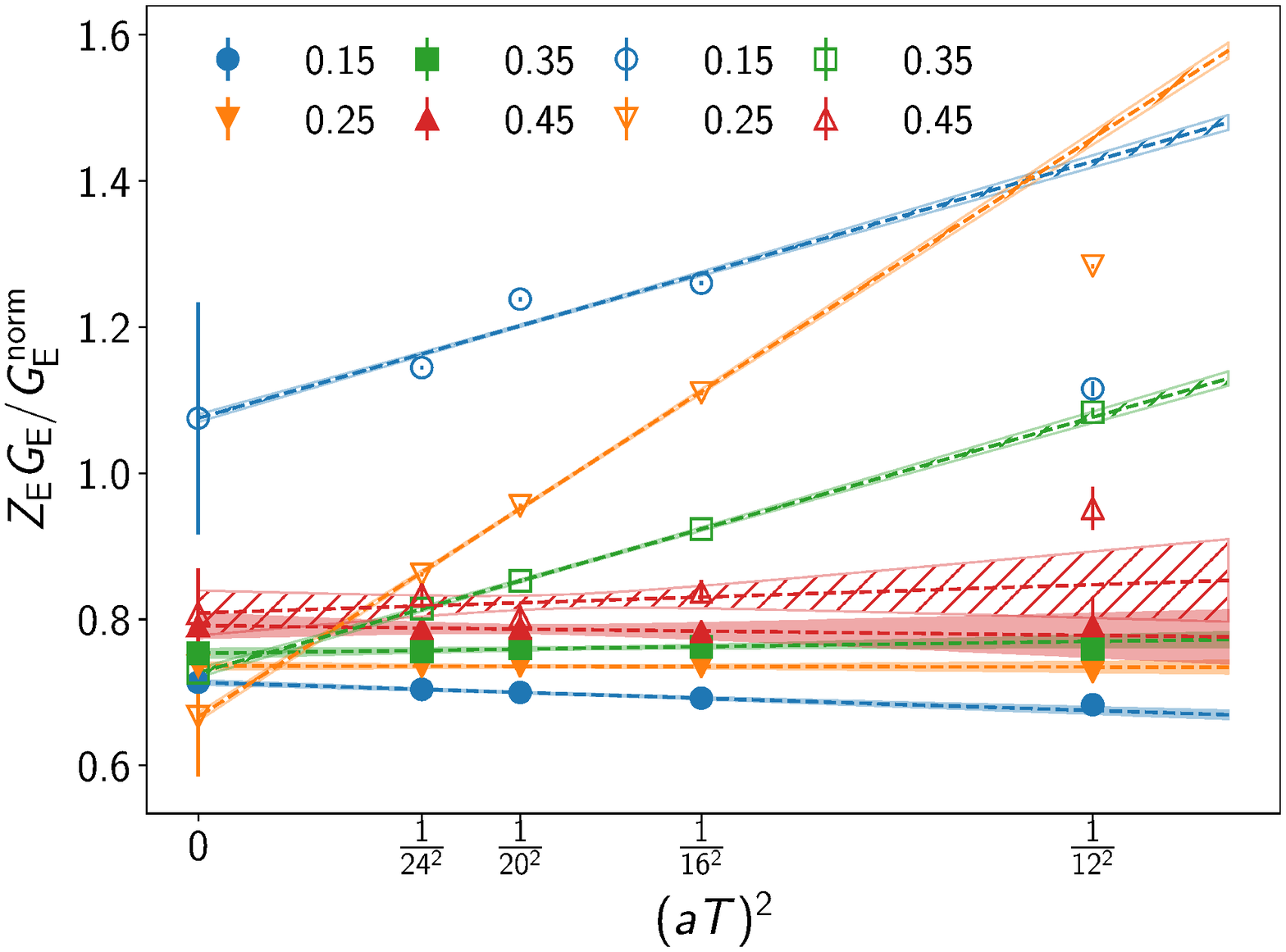}
\caption{The continuum extrapolation of the chromoelectric correlator for $T=1.5\,\Tc$ and $T=3.0\,\Tc$ (top panels)
         and for $T=6.0\,\Tc$ and $T=10^4\,\Tc$ (bottom panels). 
         The filled symbols and solid bands correspond to the extrapolation
         of the tree-level-improved lattice data, while the open symbols and patterned bands correspond to the extrapolations
         of the lattice data without tree-level improvement.
         The different colors correspond to different $\tauT$ values.}
\label{fig:contextrap_add}
\end{figure*}

\begin{figure*}
\includegraphics[width=8cm]{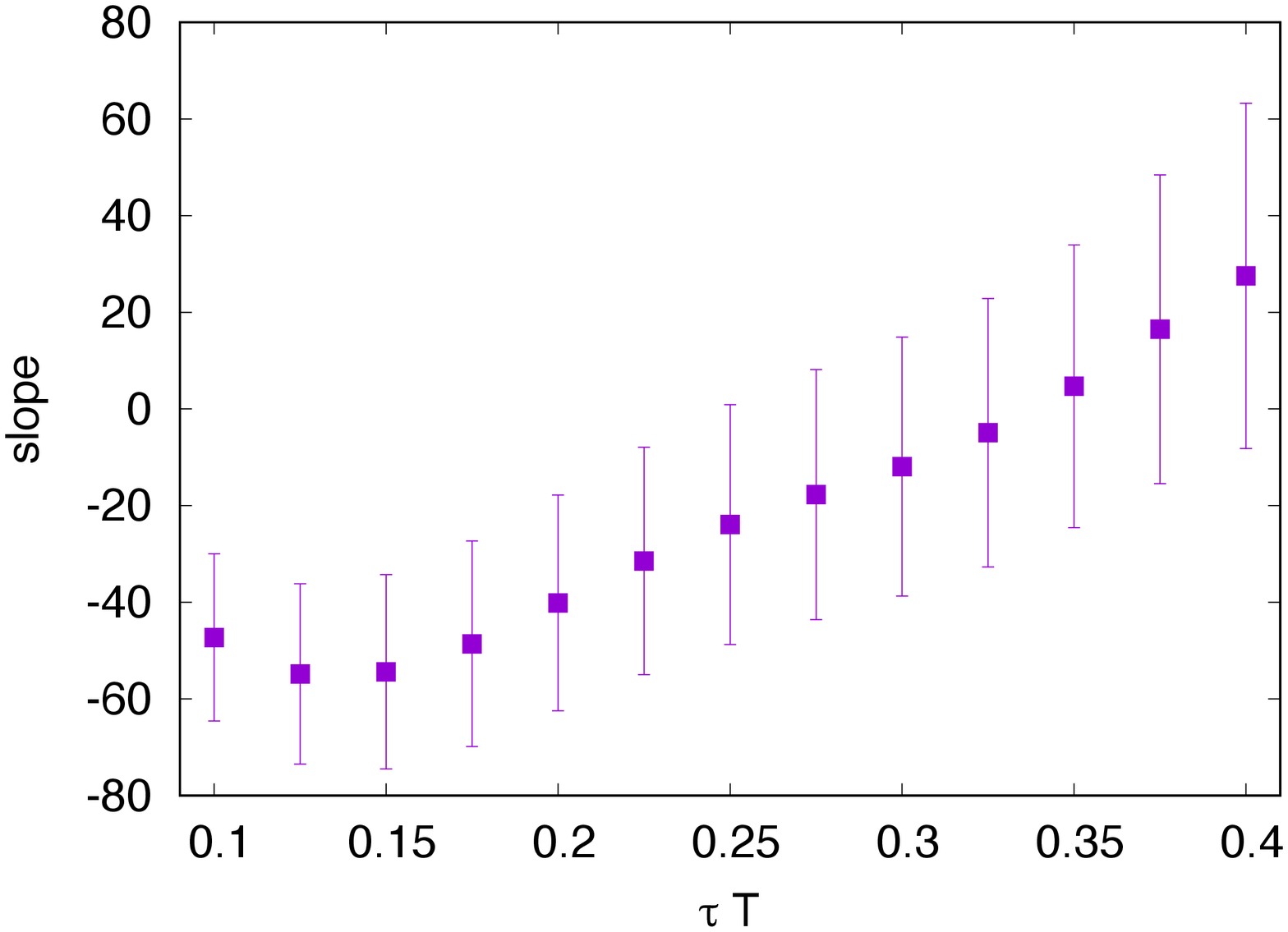}
\includegraphics[width=8cm]{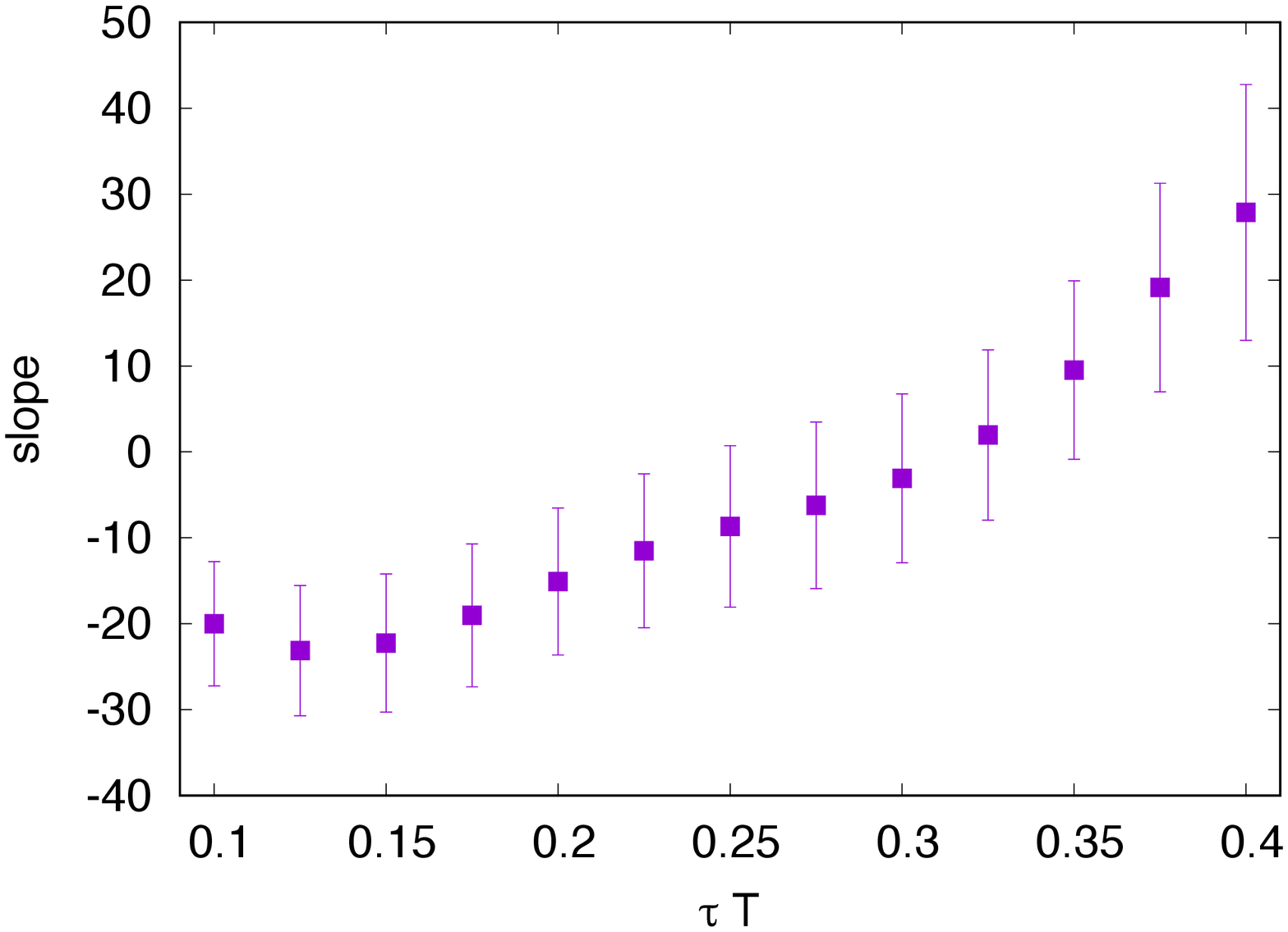}
\caption{The coefficient of the $1/\Nt^2$ dependence as a function of $\tau$ for $T=1.5\,\Tc$ (left)
and $T=10\,\Tc$ (right).}
\label{fig:atau_dep}
\end{figure*}

As discussed in the main text, to obtain the continuum result for the chromoelectric correlator, we first perform the
interpolation in $\tauT$, and then for each value of $\tauT$ we perform the continuum extrapolation using the $a/\Nt^2$
form without $\Nt=12$ data or using the $a/\Nt^2+b/\Nt^4$ form with $\Nt=12$ data included ($a$ and $b$ are fit constants). 
We have demonstrated this procedure in Fig.~\ref{fig:contextrap} for $T=1.1\,T_c$. In Fig.~\ref{fig:contextrap_add}, we show
this procedure for other temperatures: $T=1.5\,\Tc$, $3.0\,\Tc$, $6\,\Tc$, and $T=10^4\,\Tc$. We do not show the analysis 
for $10\,\Tc$, as it looks similar to the one for $T=10^4\,\Tc$. From the figure, we see that the slope
of the $1/\Nt^2$ dependence increases with decreasing $\tauT$ as expected, since the cutoff dependence is larger for
smaller $\tauT$. But for the smallest $\tauT$, we do not see this tendency. To understand the situation better, we
show the coefficient of the $1/\Nt^2$ term in the continuum extrapolation as a function of $\tauT$ in Fig.~\ref{fig:atau_dep}.
The coefficient of the $1/\Nt^2$ term decreases monotonically as $\tauT$ decreases.
For this coefficient, the minimum of the absolute value is reached at $\tauT=0.32$ instead of at the largest available $\tauT$.
Note, however, that this is somewhat accidental, as the coefficient
changes sign around this value of $\tauT$. Also, the errors for $0.3 < \tau < 0.4$ are quite large.
More importantly for us, the absolute value of the coefficient increases when we decrease $\tauT$ from 0.3 to 0.175, 
and then either flattens off or decreases if $\tauT$ is further decreased. 
We take this as an indication that the continuum limit
is not reliable for $\tauT<0.175$. 
We also perform continuum extrapolations using lattice data without tree-level improvement and the corresponding results
are also shown in Fig.~\ref{fig:contextrap_add} as open symbols. In this case, the continuum limit is always approached
from above. The continuum-extrapolated result from tree-level-improved lattice data and the unimproved lattice
data  agree within errors for $\tauT \ge 0.25$. In the absence of tree-level improvement, the continuum extrapolations
for smaller $\tauT$ are not reliable. 

\section{Modeling of the spectral function and $\kappa$ determination}\label{sec:appB}
\begin{figure*}[t]
  \includegraphics[width=8.6cm]{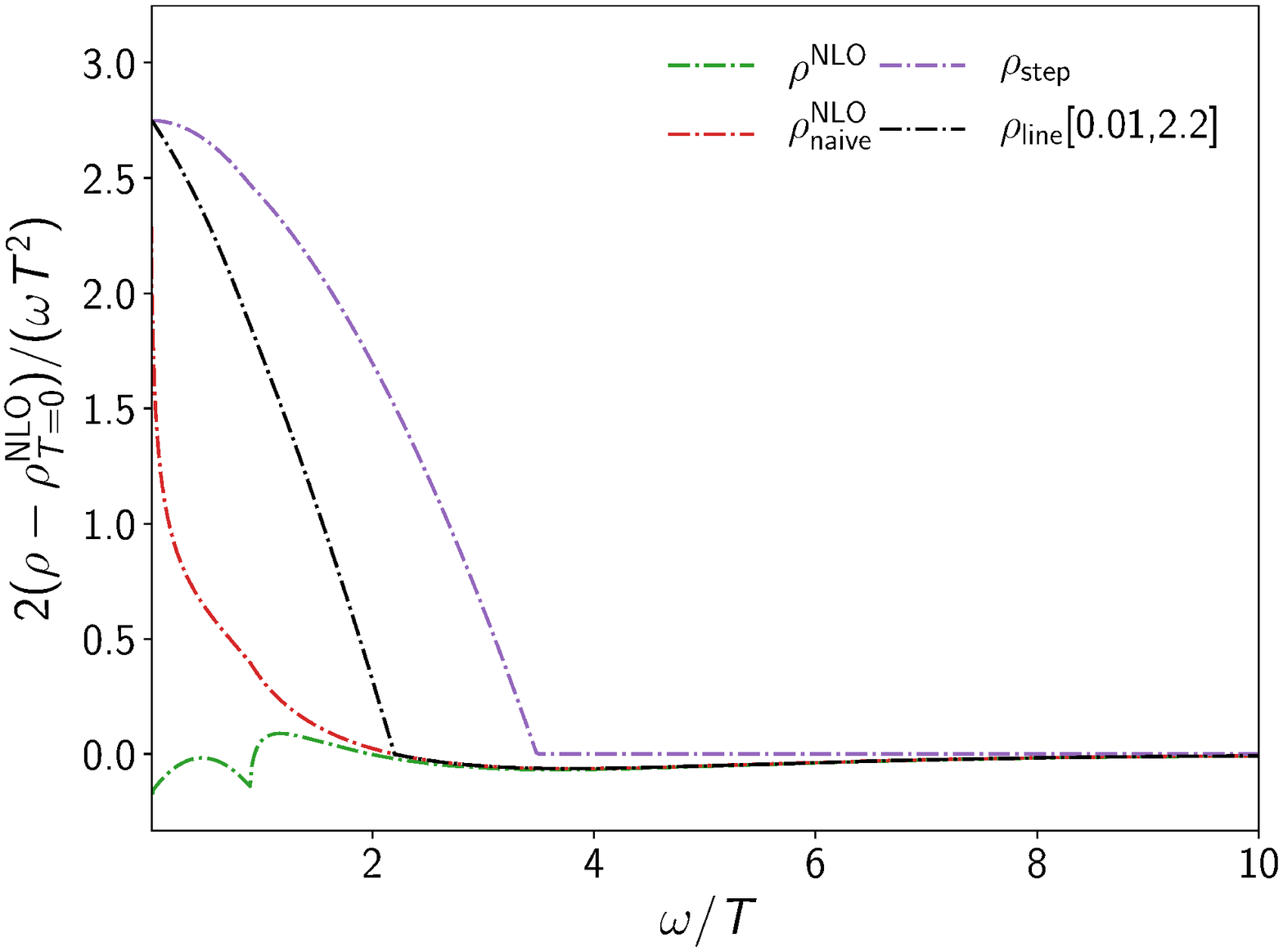}
  \includegraphics[width=8.6cm]{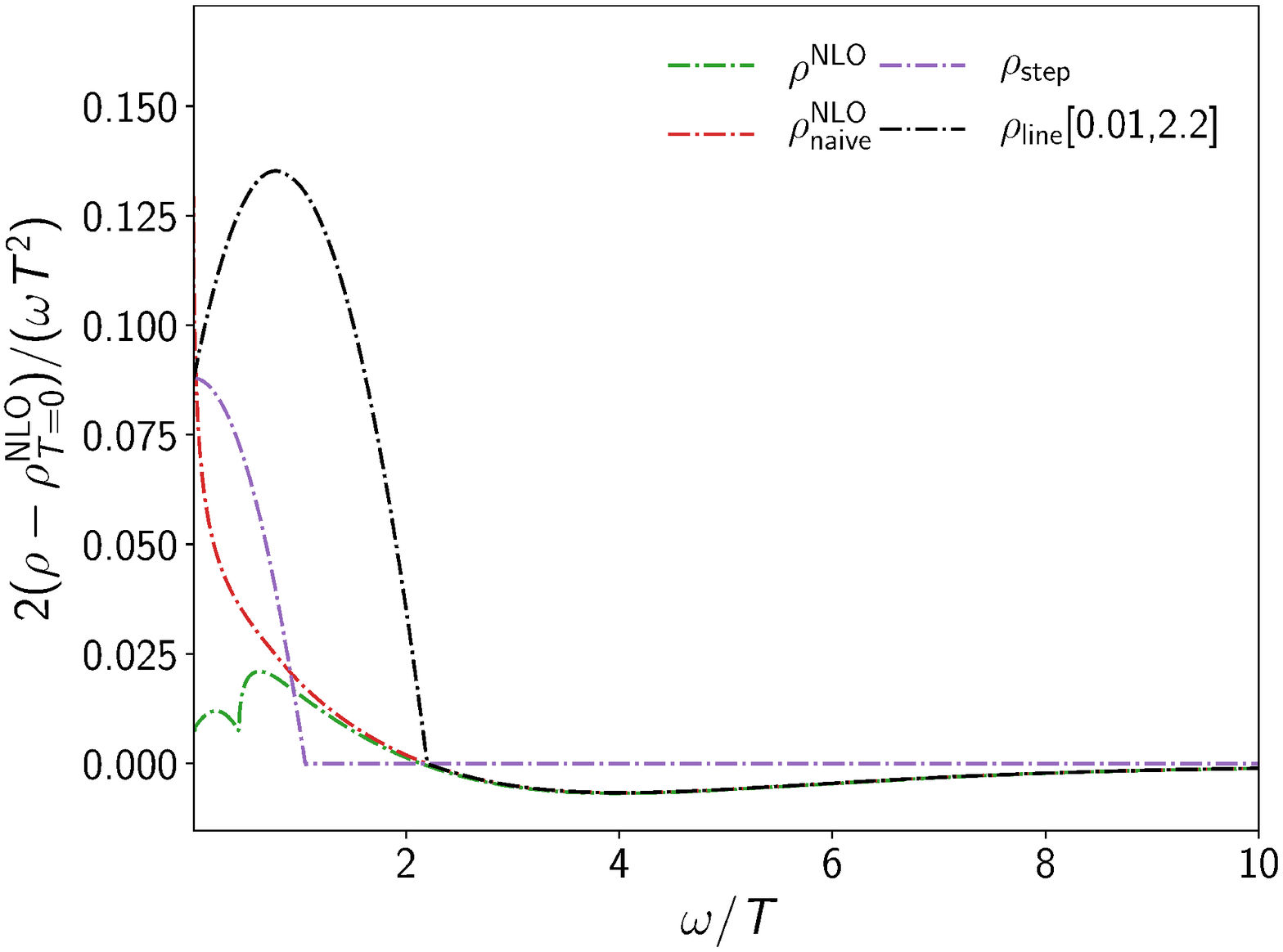} 
  \caption[b]{The NLO $T=0$ spectral function subtracted from different models or perturbative 
              curves at $T=1.1\,\Tc$ (left) and $T=10^4\,\Tc$ (right). See the text for further specifications.
  }
  \label{fig:minusspc} 
\end{figure*}

In order to understand the main features of the perturbative spectral function
corresponding to the chromoelectric correlator at NLO, in Fig. \ref{fig:minusspc} 
we show the quantity
$2(\rho(\omega,T) - \rho(\omega)_{T=0}^\mathrm{NLO})/(\omega T^2)$ calculated with
and without HTL resummation at the lowest and highest temperatures. 
The plotted quantity gives $\kappa$ in the $\omega \rightarrow 0$ limit. 
The naive (unresummed) result is logarithmically divergent at small $\omega$.
On the other hand, for $\omega>2.2\,T$, the resummed and the naive results agree well.
This indicates that the NLO calculation is valid in this $\omega$ range. 
We also see that for $2.2 < \omega/T < 10$,
the naive and resummed NLO expressions are negative, and their shapes are independent of the temperature.

In Fig.~\ref{fig:minusspc} we also show the two model spectral functions (line model and step model),
where we use $\kappa=2.75\,T^3$ for the lowest temperature and $\kappa=0.088\,T^3$ for the highest one.
At the lowest temperature, the step model has a larger finite-temperature part than
the linear model, while at the highest temperature, the opposite is true.
The two models also have somewhat different UV behavior. The step model is matched
to the zero-temperature spectral function and thus ignores the thermal correction in
the region $2.2 < \omega/T < 10$, while the line model incorporates this. 
The two models thus allow us to extract $\kappa$ 
using a set of reasonable assumptions about the large-$\omega$ behavior of the spectral function.

\begin{figure*}[t]
  \includegraphics[width=8.6cm]{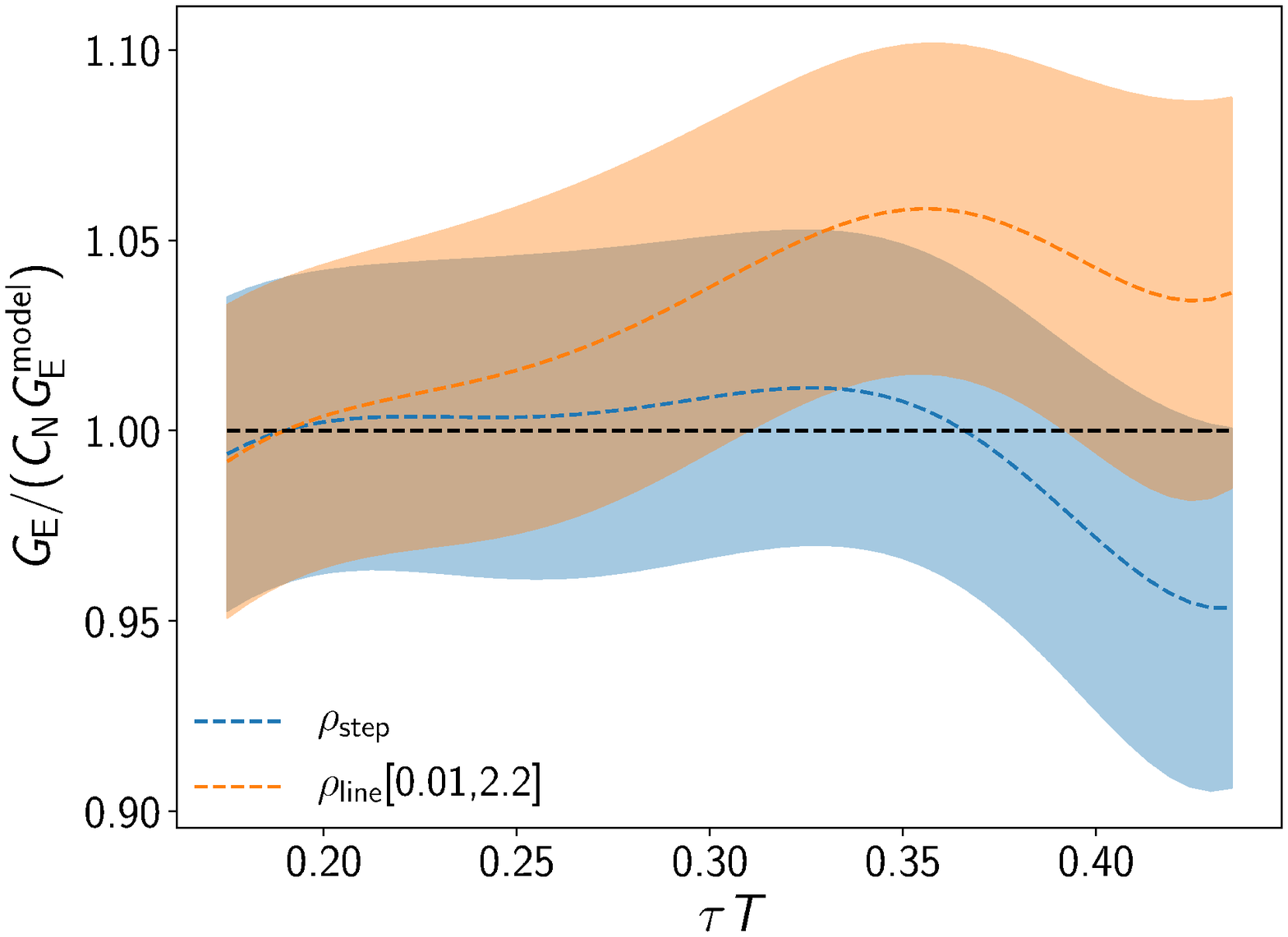}
  \includegraphics[width=8.6cm]{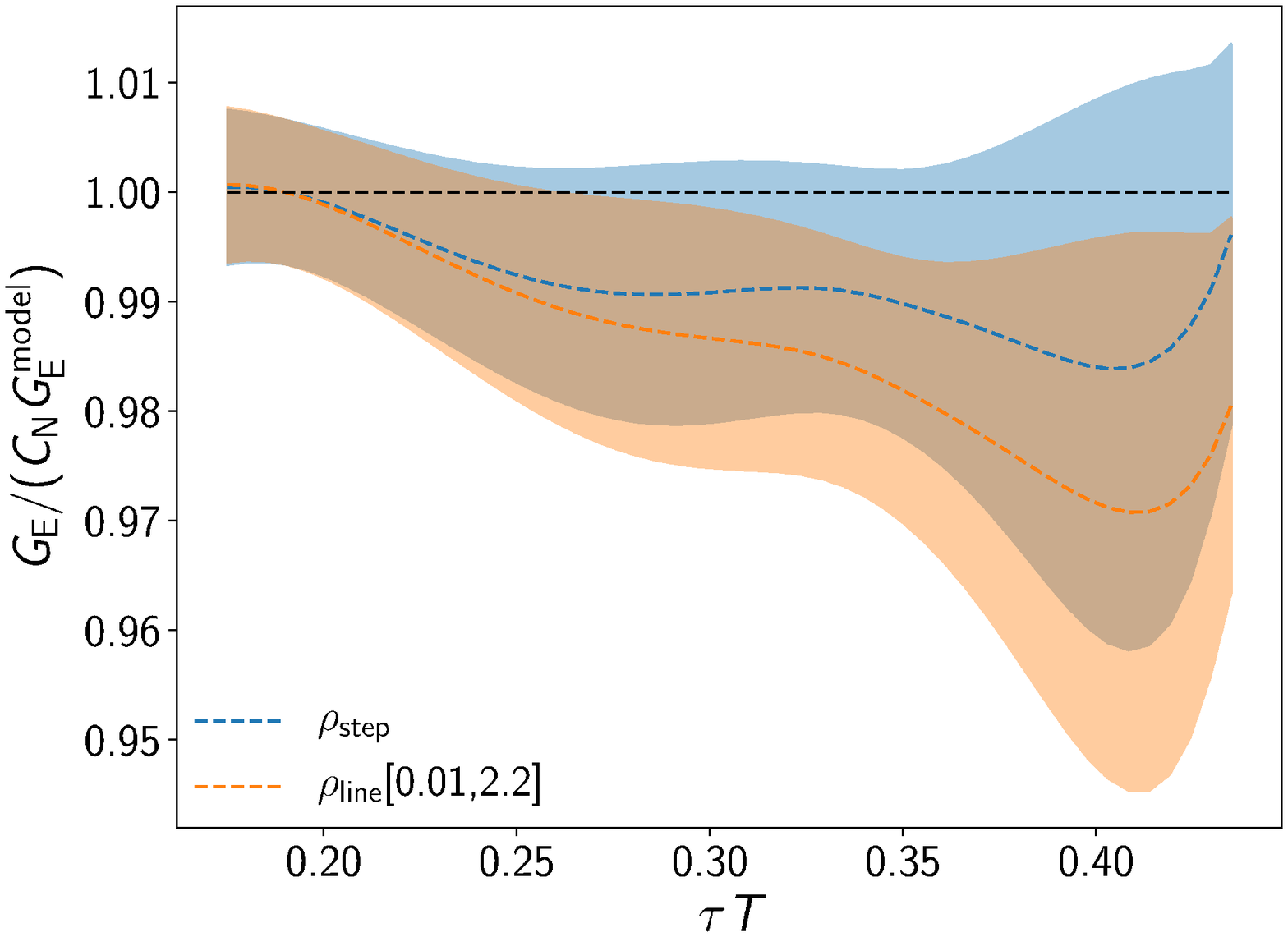}
  \caption{The ratio of the continuum-extrapolated chromoelectric correlator
           obtained on the lattice and the correlator obtained within the line model (orange band) and the step model (blue band)
           for $1.1\,\Tc$ and $\kappa = 2.75\,T^3$ (left), and for $10^4\,\Tc$ and $\kappa = 0.05\,T^3$ (right). 
           The bands show the errors originating from lattice effects and from varying the scale by a factor of 2.
  }
  \label{fig:modrat}
\end{figure*}

We match the chromoelectric correlator $\GE^\mathrm{model}$, obtained from the above model spectral
functions at $\tau T=0.19$ to the continuum-extrapolated lattice result to find the optimal
value of $\kappa$. We demonstrate this procedure in Fig.~\ref{fig:modrat} where we show the
continuum lattice result for the lowest and the highest temperatures divided by the corresponding
$\GE^\mathrm{model}$. For a given spectral function and the appropriately chosen $\kappa$ this ratio should
be close to 1. Since the errors of the continuum-extrapolated lattice result are sizable we get
a range of $\kappa$ that is compatible with the lattice result. 
In Fig.~\ref{fig:modrat} we show the results for $\kappa = 2.75\,T^3$ and
$\kappa = 0.05\,T^3$, and for $T = 1.1\,\Tc$ and $T = 10^4 \,\Tc$, respectively. 
These $\kappa$ values are chosen to be in the middle of the quoted ranges (see Eqs.~\eqref{eq:kappa11} and \eqref{eq:kappa14}).
At the lowest temperature for the given $\kappa$ the step form and the line form seem to be on the opposite side
compared to the lattice result, while at the highest temperature the step form clearly gives a better description of
the lattice data on average. We see that the optimal value of $\kappa$ strongly depends on the assumed form, especially
at high temperatures.

\FloatBarrier
\bibliography{kappa}{}
\bibliographystyle{apsrev4-1.bst}

\end{document}

%% file: table_Nconfs.dat
$T/T_c$ & $\Nt\times \Ns^3$ & $\beta$ & $N_\mathrm{conf}$ \\
\hline
&     $12 \times 48^3$  & 6.407   & 1350   \\
1.1 & $16 \times 48^3$  & 6.621   & 2623   \\
&     $20 \times 48^3$  & 6.795   & 2035   \\
&     $24 \times 48^3$  & 6.940   & 2535   \\
\hline
&     $12 \times 48^3$  & 6.639   & 1801   \\
&     $12 \times 32^3$  & 6.639   & 1557   \\
1.5 & $12 \times 24^2$  & 6.639   & 1000   \\
&     $16 \times 48^3$  & 6.872   & 2778   \\
&     $20 \times 48^3$  & 7.044   & 2081   \\
&     $24 \times 48^3$  & 7.192   & 2496   \\
\hline
2.2  & $12 \times 48^3$  & 6.940  & 1535   \\
\hline
&     $12 \times 48^3$  & 7.193   & 1579    \\
3   & $16 \times 48^3$  & 7.432   & 1553    \\
&     $20 \times 48^3$  & 7.620   & 1401    \\
&     $24 \times 48^3$  & 7.774   & 1663    \\
\hline
&     $12 \times 48^3$  & 7.774   & 1587    \\
6   & $16 \times 48^3$  & 8.019   & 1556    \\
&     $20 \times 48^3$  & 8.211   & 1258    \\
&     $24 \times 48^3$  & 8.367   & 1430    \\
\hline
&     $12 \times 48^3$  & 8.211   & 1807   \\
&     $12 \times 32^3$  & 8.211   & 1737        \\
10  & $12 \times 24^3$  & 8.211   & 1000       \\
&     $16 \times 48^3$  & 8.458   & 2769   \\
&     $20 \times 48^3$  & 8.651   & 2073   \\
&     $24 \times 48^3$  & 8.808   & 2423   \\
\hline
&     $12 \times 48^3$    & 14.194   & 1039  \\
10000 & $16 \times 48^3$  & 14.443   & 1157  \\
&     $20 \times 48^3$    & 14.635   & 1139  \\
&     $24 \times 48^3$    & 14.792   & 1375  \\
\hline
20000  & $12 \times 48^3$  & 14.792   & 1948 \\

%% file: kappa.bbl
\begin{thebibliography}{51}%
\makeatletter
\providecommand \@ifxundefined [1]{%
 \@ifx{#1\undefined}
}%
\providecommand \@ifnum [1]{%
 \ifnum #1\expandafter \@firstoftwo
 \else \expandafter \@secondoftwo
 \fi
}%
\providecommand \@ifx [1]{%
 \ifx #1\expandafter \@firstoftwo
 \else \expandafter \@secondoftwo
 \fi
}%
\providecommand \natexlab [1]{#1}%
\providecommand \enquote  [1]{``#1''}%
\providecommand \bibnamefont  [1]{#1}%
\providecommand \bibfnamefont [1]{#1}%
\providecommand \citenamefont [1]{#1}%
\providecommand \href@noop [0]{\@secondoftwo}%
\providecommand \href [0]{\begingroup \@sanitize@url \@href}%
\providecommand \@href[1]{\@@startlink{#1}\@@href}%
\providecommand \@@href[1]{\endgroup#1\@@endlink}%
\providecommand \@sanitize@url [0]{\catcode `\\12\catcode `\$12\catcode
  `\&12\catcode `\#12\catcode `\^12\catcode `\_12\catcode `\%12\relax}%
\providecommand \@@startlink[1]{}%
\providecommand \@@endlink[0]{}%
\providecommand \url  [0]{\begingroup\@sanitize@url \@url }%
\providecommand \@url [1]{\endgroup\@href {#1}{\urlprefix }}%
\providecommand \urlprefix  [0]{URL }%
\providecommand \Eprint [0]{\href }%
\@ifxundefined \urlstyle {%
  \providecommand \doi  [0]{\begingroup \@sanitize@url \@doi}%
  \providecommand \@doi [1]{\endgroup \@@startlink {\doibase
  #1}doi:\discretionary {}{}{}#1\@@endlink }%
}{%
  \providecommand \doi  [0]{doi:\discretionary{}{}{}\begingroup
  \urlstyle{rm}\Url }%
}%
\providecommand \doibase [0]{http://dx.doi.org/}%
\providecommand \Doi [0]{\begingroup \@sanitize@url \@Doi }%
\providecommand \@Doi  [1]{\endgroup\@@startlink{\doibase#1}\@@Doi}%
\providecommand \@@Doi [1]{#1\@@endlink}%
\providecommand \selectlanguage [0]{\@gobble}%
\providecommand \bibinfo  [0]{\@secondoftwo}%
\providecommand \bibfield  [0]{\@secondoftwo}%
\providecommand \translation [1]{[#1]}%
\providecommand \BibitemOpen [0]{}%
\providecommand \bibitemStop [0]{}%
\providecommand \bibitemNoStop [0]{.\EOS\space}%
\providecommand \EOS [0]{\spacefactor3000\relax}%
\providecommand \BibitemShut  [1]{\csname bibitem#1\endcsname}%
\bibitem [{\citenamefont {Busza}\ \emph {et~al.}(2018)\citenamefont {Busza},
  \citenamefont {Rajagopal},\ and\ \citenamefont {van~der
  Schee}}]{Busza:2018rrf}%
  \BibitemOpen
  \bibfield  {author} {\bibinfo {author} {\bibfnamefont {W.}~\bibnamefont
  {Busza}}, \bibinfo {author} {\bibfnamefont {K.}~\bibnamefont {Rajagopal}}, \
  and\ \bibinfo {author} {\bibfnamefont {W.}~\bibnamefont {van~der Schee}},\
  }\Doi {10.1146/annurev-nucl-101917-020852} {\bibfield  {journal} {\bibinfo
  {journal} {Ann. Rev. Nucl. Part. Sci.},\ }\textbf {\bibinfo {volume} {68}},\
  \bibinfo {pages} {339} (\bibinfo {year} {2018})},\ \Eprint
  {http://arxiv.org/abs/1802.04801} {arXiv:1802.04801 [hep-ph]} \BibitemShut
  {NoStop}%
\bibitem [{\citenamefont {Shuryak}\ and\ \citenamefont
  {Zahed}(2004)}]{Shuryak:2003ty}%
  \BibitemOpen
  \bibfield  {author} {\bibinfo {author} {\bibfnamefont {E.~V.}\ \bibnamefont
  {Shuryak}}\ and\ \bibinfo {author} {\bibfnamefont {I.}~\bibnamefont
  {Zahed}},\ }\Doi {10.1103/PhysRevC.70.021901} {\bibfield  {journal} {\bibinfo
   {journal} {Phys. Rev.},\ }\textbf {\bibinfo {volume} {C70}},\ \bibinfo
  {pages} {021901} (\bibinfo {year} {2004})},\ \Eprint
  {http://arxiv.org/abs/hep-ph/0307267} {arXiv:hep-ph/0307267 [hep-ph]}
  \BibitemShut {NoStop}%
\bibitem [{\citenamefont {Shuryak}(2004)}]{Shuryak:2003xe}%
  \BibitemOpen
  \bibfield  {author} {\bibinfo {author} {\bibfnamefont {E.}~\bibnamefont
  {Shuryak}},\ }\bibfield  {booktitle} {\emph {\bibinfo {booktitle} {{Heavy ion
  reaction from nuclear to quark matter. Proceedings, International School of
  Nuclear Physics, 25th Course, Erice, Italy, September 16-24, 2003}}},\ }\Doi
  {10.1016/j.ppnp.2004.02.025} {\bibfield  {journal} {\bibinfo  {journal}
  {Prog. Part. Nucl. Phys.},\ }\textbf {\bibinfo {volume} {53}},\ \bibinfo
  {pages} {273} (\bibinfo {year} {2004})},\ \Eprint
  {http://arxiv.org/abs/hep-ph/0312227} {arXiv:hep-ph/0312227 [hep-ph]}
  \BibitemShut {NoStop}%
\bibitem [{\citenamefont {Beraudo}\ \emph {et~al.}(2018)\citenamefont {Beraudo}
  \emph {et~al.}}]{Rapp:2018qla}%
  \BibitemOpen
  \bibfield  {author} {\bibinfo {author} {\bibfnamefont {A.}~\bibnamefont
  {Beraudo}} \emph {et~al.},\ }\Doi {10.1016/j.nuclphysa.2018.09.002}
  {\bibfield  {journal} {\bibinfo  {journal} {Nucl. Phys.},\ }\textbf {\bibinfo
  {volume} {A979}},\ \bibinfo {pages} {21} (\bibinfo {year} {2018})},\ \Eprint
  {http://arxiv.org/abs/1803.03824} {arXiv:1803.03824 [nucl-th]} \BibitemShut
  {NoStop}%
\bibitem [{\citenamefont {Moore}\ and\ \citenamefont
  {Teaney}(2005)}]{Moore:2004tg}%
  \BibitemOpen
  \bibfield  {author} {\bibinfo {author} {\bibfnamefont {G.~D.}\ \bibnamefont
  {Moore}}\ and\ \bibinfo {author} {\bibfnamefont {D.}~\bibnamefont {Teaney}},\
  }\Doi {10.1103/PhysRevC.71.064904} {\bibfield  {journal} {\bibinfo  {journal}
  {Phys. Rev.},\ }\textbf {\bibinfo {volume} {C71}},\ \bibinfo {pages} {064904}
  (\bibinfo {year} {2005})},\ \Eprint {http://arxiv.org/abs/hep-ph/0412346}
  {arXiv:hep-ph/0412346 [hep-ph]} \BibitemShut {NoStop}%
\bibitem [{\citenamefont {Svetitsky}(1988)}]{Svetitsky:1987gq}%
  \BibitemOpen
  \bibfield  {author} {\bibinfo {author} {\bibfnamefont {B.}~\bibnamefont
  {Svetitsky}},\ }\Doi {10.1103/PhysRevD.37.2484} {\bibfield  {journal}
  {\bibinfo  {journal} {Phys. Rev.},\ }\textbf {\bibinfo {volume} {D37}},\
  \bibinfo {pages} {2484} (\bibinfo {year} {1988})}\BibitemShut {NoStop}%
\bibitem [{\citenamefont {Caron-Huot}\ and\ \citenamefont
  {Moore}(2008){\natexlab{a}}}]{CaronHuot:2008uh}%
  \BibitemOpen
  \bibfield  {author} {\bibinfo {author} {\bibfnamefont {S.}~\bibnamefont
  {Caron-Huot}}\ and\ \bibinfo {author} {\bibfnamefont {G.~D.}\ \bibnamefont
  {Moore}},\ }\Doi {10.1088/1126-6708/2008/02/081} {\bibfield  {journal}
  {\bibinfo  {journal} {JHEP},\ }\textbf {\bibinfo {volume} {02}},\ \bibinfo
  {pages} {081} (\bibinfo {year} {2008}{\natexlab{a}})},\ \Eprint
  {http://arxiv.org/abs/0801.2173} {arXiv:0801.2173 [hep-ph]} \BibitemShut
  {NoStop}%
\bibitem [{\citenamefont {Herzog}\ \emph {et~al.}(2006)\citenamefont {Herzog},
  \citenamefont {Karch}, \citenamefont {Kovtun}, \citenamefont {Kozcaz},\ and\
  \citenamefont {Yaffe}}]{Herzog:2006gh}%
  \BibitemOpen
  \bibfield  {author} {\bibinfo {author} {\bibfnamefont {C.~P.}\ \bibnamefont
  {Herzog}}, \bibinfo {author} {\bibfnamefont {A.}~\bibnamefont {Karch}},
  \bibinfo {author} {\bibfnamefont {P.}~\bibnamefont {Kovtun}}, \bibinfo
  {author} {\bibfnamefont {C.}~\bibnamefont {Kozcaz}}, \ and\ \bibinfo {author}
  {\bibfnamefont {L.~G.}\ \bibnamefont {Yaffe}},\ }\Doi
  {10.1088/1126-6708/2006/07/013} {\bibfield  {journal} {\bibinfo  {journal}
  {JHEP},\ }\textbf {\bibinfo {volume} {07}},\ \bibinfo {pages} {013} (\bibinfo
  {year} {2006})},\ \Eprint {http://arxiv.org/abs/hep-th/0605158}
  {arXiv:hep-th/0605158 [hep-th]} \BibitemShut {NoStop}%
\bibitem [{\citenamefont {Casalderrey-Solana}\ and\ \citenamefont
  {Teaney}(2006)}]{CasalderreySolana:2006rq}%
  \BibitemOpen
  \bibfield  {author} {\bibinfo {author} {\bibfnamefont {J.}~\bibnamefont
  {Casalderrey-Solana}}\ and\ \bibinfo {author} {\bibfnamefont
  {D.}~\bibnamefont {Teaney}},\ }\Doi {10.1103/PhysRevD.74.085012} {\bibfield
  {journal} {\bibinfo  {journal} {Phys. Rev.},\ }\textbf {\bibinfo {volume}
  {D74}},\ \bibinfo {pages} {085012} (\bibinfo {year} {2006})},\ \Eprint
  {http://arxiv.org/abs/hep-ph/0605199} {arXiv:hep-ph/0605199 [hep-ph]}
  \BibitemShut {NoStop}%
\bibitem [{\citenamefont {Aarts}\ and\ \citenamefont
  {Martinez~Resco}(2002)}]{Aarts:2002cc}%
  \BibitemOpen
  \bibfield  {author} {\bibinfo {author} {\bibfnamefont {G.}~\bibnamefont
  {Aarts}}\ and\ \bibinfo {author} {\bibfnamefont {J.~M.}\ \bibnamefont
  {Martinez~Resco}},\ }\Doi {10.1088/1126-6708/2002/04/053} {\bibfield
  {journal} {\bibinfo  {journal} {JHEP},\ }\textbf {\bibinfo {volume} {04}},\
  \bibinfo {pages} {053} (\bibinfo {year} {2002})},\ \Eprint
  {http://arxiv.org/abs/hep-ph/0203177} {arXiv:hep-ph/0203177 [hep-ph]}
  \BibitemShut {NoStop}%
\bibitem [{\citenamefont {Petreczky}\ and\ \citenamefont
  {Teaney}(2006)}]{Petreczky:2005nh}%
  \BibitemOpen
  \bibfield  {author} {\bibinfo {author} {\bibfnamefont {P.}~\bibnamefont
  {Petreczky}}\ and\ \bibinfo {author} {\bibfnamefont {D.}~\bibnamefont
  {Teaney}},\ }\Doi {10.1103/PhysRevD.73.014508} {\bibfield  {journal}
  {\bibinfo  {journal} {Phys. Rev.},\ }\textbf {\bibinfo {volume} {D73}},\
  \bibinfo {pages} {014508} (\bibinfo {year} {2006})},\ \Eprint
  {http://arxiv.org/abs/hep-ph/0507318} {arXiv:hep-ph/0507318 [hep-ph]}
  \BibitemShut {NoStop}%
\bibitem [{\citenamefont {Petreczky}(2009)}]{Petreczky:2008px}%
  \BibitemOpen
  \bibfield  {author} {\bibinfo {author} {\bibfnamefont {P.}~\bibnamefont
  {Petreczky}},\ }\Doi {10.1140/epjc/s10052-009-0942-1} {\bibfield  {journal}
  {\bibinfo  {journal} {Eur. Phys. J. C},\ }\textbf {\bibinfo {volume} {62}},\
  \bibinfo {pages} {85} (\bibinfo {year} {2009})},\ \Eprint
  {http://arxiv.org/abs/0810.0258} {arXiv:0810.0258 [hep-lat]} \BibitemShut
  {NoStop}%
\bibitem [{\citenamefont {Ding}\ \emph {et~al.}(2012)\citenamefont {Ding},
  \citenamefont {Francis}, \citenamefont {Kaczmarek}, \citenamefont {Karsch},
  \citenamefont {Satz},\ and\ \citenamefont {Soeldner}}]{Ding:2012sp}%
  \BibitemOpen
  \bibfield  {author} {\bibinfo {author} {\bibfnamefont {H.~T.}\ \bibnamefont
  {Ding}}, \bibinfo {author} {\bibfnamefont {A.}~\bibnamefont {Francis}},
  \bibinfo {author} {\bibfnamefont {O.}~\bibnamefont {Kaczmarek}}, \bibinfo
  {author} {\bibfnamefont {F.}~\bibnamefont {Karsch}}, \bibinfo {author}
  {\bibfnamefont {H.}~\bibnamefont {Satz}}, \ and\ \bibinfo {author}
  {\bibfnamefont {W.}~\bibnamefont {Soeldner}},\ }\Doi
  {10.1103/PhysRevD.86.014509} {\bibfield  {journal} {\bibinfo  {journal}
  {Phys. Rev.},\ }\textbf {\bibinfo {volume} {D86}},\ \bibinfo {pages} {014509}
  (\bibinfo {year} {2012})},\ \Eprint {http://arxiv.org/abs/1204.4945}
  {arXiv:1204.4945 [hep-lat]} \BibitemShut {NoStop}%
\bibitem [{\citenamefont {Ding}\ \emph {et~al.}(2019)\citenamefont {Ding},
  \citenamefont {Kaczmarek}, \citenamefont {Kruse}, \citenamefont {Larsen},
  \citenamefont {Mazur}, \citenamefont {Mukherjee}, \citenamefont {Ohno},
  \citenamefont {Sandmeyer},\ and\ \citenamefont {Shu}}]{Ding:2018uhl}%
  \BibitemOpen
  \bibfield  {author} {\bibinfo {author} {\bibfnamefont {H.-T.}\ \bibnamefont
  {Ding}}, \bibinfo {author} {\bibfnamefont {O.}~\bibnamefont {Kaczmarek}},
  \bibinfo {author} {\bibfnamefont {A.-L.}\ \bibnamefont {Kruse}}, \bibinfo
  {author} {\bibfnamefont {R.}~\bibnamefont {Larsen}}, \bibinfo {author}
  {\bibfnamefont {L.}~\bibnamefont {Mazur}}, \bibinfo {author} {\bibfnamefont
  {S.}~\bibnamefont {Mukherjee}}, \bibinfo {author} {\bibfnamefont
  {H.}~\bibnamefont {Ohno}}, \bibinfo {author} {\bibfnamefont {H.}~\bibnamefont
  {Sandmeyer}}, \ and\ \bibinfo {author} {\bibfnamefont {H.-T.}\ \bibnamefont
  {Shu}},\ }\bibfield  {booktitle} {\emph {\bibinfo {booktitle} {{Proceedings,
  27th International Conference on Ultrarelativistic Nucleus-Nucleus Collisions
  (Quark Matter 2018): Venice, Italy, May 14-19, 2018}}},\ }\Doi
  {10.1016/j.nuclphysa.2018.09.075} {\bibfield  {journal} {\bibinfo  {journal}
  {Nucl. Phys.},\ }\textbf {\bibinfo {volume} {A982}},\ \bibinfo {pages} {715}
  (\bibinfo {year} {2019})},\ \Eprint {http://arxiv.org/abs/1807.06315}
  {arXiv:1807.06315 [hep-lat]} \BibitemShut {NoStop}%
\bibitem [{\citenamefont {Lorenz}\ \emph {et~al.}(2020)\citenamefont {Lorenz},
  \citenamefont {Ding}, \citenamefont {Kaczmarek}, \citenamefont {Ohno},
  \citenamefont {Sandmeyer},\ and\ \citenamefont {Shu}}]{Lorenz:2020uik}%
  \BibitemOpen
  \bibfield  {author} {\bibinfo {author} {\bibfnamefont {A.-L.}\ \bibnamefont
  {Lorenz}}, \bibinfo {author} {\bibfnamefont {H.-T.}\ \bibnamefont {Ding}},
  \bibinfo {author} {\bibfnamefont {O.}~\bibnamefont {Kaczmarek}}, \bibinfo
  {author} {\bibfnamefont {H.}~\bibnamefont {Ohno}}, \bibinfo {author}
  {\bibfnamefont {H.}~\bibnamefont {Sandmeyer}}, \ and\ \bibinfo {author}
  {\bibfnamefont {H.-T.}\ \bibnamefont {Shu}},\ }\Doi {10.22323/1.363.0207}
  {\bibfield  {journal} {\bibinfo  {journal} {PoS},\ }\textbf {\bibinfo
  {volume} {LATTICE2019}},\ \bibinfo {pages} {207} (\bibinfo {year} {2020})},\
  \Eprint {http://arxiv.org/abs/2002.00681} {arXiv:2002.00681 [hep-lat]}
  \BibitemShut {NoStop}%
\bibitem [{\citenamefont {Borsanyi}\ \emph {et~al.}(2014)\citenamefont
  {Borsanyi} \emph {et~al.}}]{Borsanyi:2014vka}%
  \BibitemOpen
  \bibfield  {author} {\bibinfo {author} {\bibfnamefont {S.}~\bibnamefont
  {Borsanyi}} \emph {et~al.},\ }\Doi {10.1007/JHEP04(2014)132} {\bibfield
  {journal} {\bibinfo  {journal} {JHEP},\ }\textbf {\bibinfo {volume} {04}},\
  \bibinfo {pages} {132} (\bibinfo {year} {2014})},\ \Eprint
  {http://arxiv.org/abs/1401.5940} {arXiv:1401.5940 [hep-lat]} \BibitemShut
  {NoStop}%
\bibitem [{\citenamefont {Boguslavski}\ \emph
  {et~al.}(2020){\natexlab{a}}\citenamefont {Boguslavski}, \citenamefont
  {Kurkela}, \citenamefont {Lappi},\ and\ \citenamefont
  {Peuron}}]{Boguslavski:2020mzh}%
  \BibitemOpen
  \bibfield  {author} {\bibinfo {author} {\bibfnamefont {K.}~\bibnamefont
  {Boguslavski}}, \bibinfo {author} {\bibfnamefont {A.}~\bibnamefont
  {Kurkela}}, \bibinfo {author} {\bibfnamefont {T.}~\bibnamefont {Lappi}}, \
  and\ \bibinfo {author} {\bibfnamefont {J.}~\bibnamefont {Peuron}},\ }in\
  \href@noop {} {\emph {\bibinfo {booktitle} {{28th International Conference on
  Ultrarelativistic Nucleus-Nucleus Collisions}}}}\ (\bibinfo {year} {2020})\
  \Eprint {http://arxiv.org/abs/2001.11863} {arXiv:2001.11863 [hep-ph]}
  \BibitemShut {NoStop}%
\bibitem [{\citenamefont {Boguslavski}\ \emph
  {et~al.}(2020){\natexlab{b}}\citenamefont {Boguslavski}, \citenamefont
  {Kurkela}, \citenamefont {Lappi},\ and\ \citenamefont
  {Peuron}}]{Boguslavski:2020tqz}%
  \BibitemOpen
  \bibfield  {author} {\bibinfo {author} {\bibfnamefont {K.}~\bibnamefont
  {Boguslavski}}, \bibinfo {author} {\bibfnamefont {A.}~\bibnamefont
  {Kurkela}}, \bibinfo {author} {\bibfnamefont {T.}~\bibnamefont {Lappi}}, \
  and\ \bibinfo {author} {\bibfnamefont {J.}~\bibnamefont {Peuron}},\ }\Doi
  {10.1007/JHEP09(2020)077} {\bibfield  {journal} {\bibinfo  {journal} {JHEP},\
  }\textbf {\bibinfo {volume} {09}},\ \bibinfo {pages} {077} (\bibinfo {year}
  {2020}{\natexlab{b}})},\ \Eprint {http://arxiv.org/abs/2005.02418}
  {arXiv:2005.02418 [hep-ph]} \BibitemShut {NoStop}%
\bibitem [{\citenamefont {Brambilla}\ \emph {et~al.}(2017)\citenamefont
  {Brambilla}, \citenamefont {Escobedo}, \citenamefont {Soto},\ and\
  \citenamefont {Vairo}}]{Brambilla:2016wgg}%
  \BibitemOpen
  \bibfield  {author} {\bibinfo {author} {\bibfnamefont {N.}~\bibnamefont
  {Brambilla}}, \bibinfo {author} {\bibfnamefont {M.~A.}\ \bibnamefont
  {Escobedo}}, \bibinfo {author} {\bibfnamefont {J.}~\bibnamefont {Soto}}, \
  and\ \bibinfo {author} {\bibfnamefont {A.}~\bibnamefont {Vairo}},\ }\Doi
  {10.1103/PhysRevD.96.034021} {\bibfield  {journal} {\bibinfo  {journal}
  {Phys. Rev.},\ }\textbf {\bibinfo {volume} {D96}},\ \bibinfo {pages} {034021}
  (\bibinfo {year} {2017})},\ \Eprint {http://arxiv.org/abs/1612.07248}
  {arXiv:1612.07248 [hep-ph]} \BibitemShut {NoStop}%
\bibitem [{\citenamefont {Brambilla}\ \emph {et~al.}(2018)\citenamefont
  {Brambilla}, \citenamefont {Escobedo}, \citenamefont {Soto},\ and\
  \citenamefont {Vairo}}]{Brambilla:2017zei}%
  \BibitemOpen
  \bibfield  {author} {\bibinfo {author} {\bibfnamefont {N.}~\bibnamefont
  {Brambilla}}, \bibinfo {author} {\bibfnamefont {M.~A.}\ \bibnamefont
  {Escobedo}}, \bibinfo {author} {\bibfnamefont {J.}~\bibnamefont {Soto}}, \
  and\ \bibinfo {author} {\bibfnamefont {A.}~\bibnamefont {Vairo}},\ }\Doi
  {10.1103/PhysRevD.97.074009} {\bibfield  {journal} {\bibinfo  {journal}
  {Phys. Rev.},\ }\textbf {\bibinfo {volume} {D97}},\ \bibinfo {pages} {074009}
  (\bibinfo {year} {2018})},\ \Eprint {http://arxiv.org/abs/1711.04515}
  {arXiv:1711.04515 [hep-ph]} \BibitemShut {NoStop}%
\bibitem [{\citenamefont {Brambilla}\ \emph {et~al.}(2019)\citenamefont
  {Brambilla}, \citenamefont {Escobedo}, \citenamefont {Vairo},\ and\
  \citenamefont {Vander~Griend}}]{Brambilla:2019tpt}%
  \BibitemOpen
  \bibfield  {author} {\bibinfo {author} {\bibfnamefont {N.}~\bibnamefont
  {Brambilla}}, \bibinfo {author} {\bibfnamefont {M.~A.}\ \bibnamefont
  {Escobedo}}, \bibinfo {author} {\bibfnamefont {A.}~\bibnamefont {Vairo}}, \
  and\ \bibinfo {author} {\bibfnamefont {P.}~\bibnamefont {Vander~Griend}},\
  }\Doi {10.1103/PhysRevD.100.054025} {\bibfield  {journal} {\bibinfo
  {journal} {Phys. Rev.},\ }\textbf {\bibinfo {volume} {D100}},\ \bibinfo
  {pages} {054025} (\bibinfo {year} {2019})},\ \Eprint
  {http://arxiv.org/abs/1903.08063} {arXiv:1903.08063 [hep-ph]} \BibitemShut
  {NoStop}%
\bibitem [{\citenamefont {Caron-Huot}\ \emph {et~al.}(2009)\citenamefont
  {Caron-Huot}, \citenamefont {Laine},\ and\ \citenamefont
  {Moore}}]{CaronHuot:2009uh}%
  \BibitemOpen
  \bibfield  {author} {\bibinfo {author} {\bibfnamefont {S.}~\bibnamefont
  {Caron-Huot}}, \bibinfo {author} {\bibfnamefont {M.}~\bibnamefont {Laine}}, \
  and\ \bibinfo {author} {\bibfnamefont {G.~D.}\ \bibnamefont {Moore}},\ }\Doi
  {10.1088/1126-6708/2009/04/053} {\bibfield  {journal} {\bibinfo  {journal}
  {JHEP},\ }\textbf {\bibinfo {volume} {04}},\ \bibinfo {pages} {053} (\bibinfo
  {year} {2009})},\ \Eprint {http://arxiv.org/abs/0901.1195} {arXiv:0901.1195
  [hep-lat]} \BibitemShut {NoStop}%
\bibitem [{\citenamefont {Meyer}(2011)}]{Meyer:2010tt}%
  \BibitemOpen
  \bibfield  {author} {\bibinfo {author} {\bibfnamefont {H.~B.}\ \bibnamefont
  {Meyer}},\ }\Doi {10.1088/1367-2630/13/3/035008} {\bibfield  {journal}
  {\bibinfo  {journal} {New J. Phys.},\ }\textbf {\bibinfo {volume} {13}},\
  \bibinfo {pages} {035008} (\bibinfo {year} {2011})},\ \Eprint
  {http://arxiv.org/abs/1012.0234} {arXiv:1012.0234 [hep-lat]} \BibitemShut
  {NoStop}%
\bibitem [{\citenamefont {Francis}\ \emph {et~al.}(2011)\citenamefont
  {Francis}, \citenamefont {Kaczmarek}, \citenamefont {Laine},\ and\
  \citenamefont {Langelage}}]{Francis:2011gc}%
  \BibitemOpen
  \bibfield  {author} {\bibinfo {author} {\bibfnamefont {A.}~\bibnamefont
  {Francis}}, \bibinfo {author} {\bibfnamefont {O.}~\bibnamefont {Kaczmarek}},
  \bibinfo {author} {\bibfnamefont {M.}~\bibnamefont {Laine}}, \ and\ \bibinfo
  {author} {\bibfnamefont {J.}~\bibnamefont {Langelage}},\ }\bibfield
  {booktitle} {\emph {\bibinfo {booktitle} {{Proceedings, 29th International
  Symposium on Lattice field theory (Lattice 2011): Squaw Valley, Lake Tahoe,
  USA, July 10-16, 2011}}},\ }\Doi {10.22323/1.139.0202} {\bibfield  {journal}
  {\bibinfo  {journal} {PoS},\ }\textbf {\bibinfo {volume} {LATTICE2011}},\
  \bibinfo {pages} {202} (\bibinfo {year} {2011})},\ \Eprint
  {http://arxiv.org/abs/1109.3941} {arXiv:1109.3941 [hep-lat]} \BibitemShut
  {NoStop}%
\bibitem [{\citenamefont {Banerjee}\ \emph {et~al.}(2012)\citenamefont
  {Banerjee}, \citenamefont {Datta}, \citenamefont {Gavai},\ and\ \citenamefont
  {Majumdar}}]{Banerjee:2011ra}%
  \BibitemOpen
  \bibfield  {author} {\bibinfo {author} {\bibfnamefont {D.}~\bibnamefont
  {Banerjee}}, \bibinfo {author} {\bibfnamefont {S.}~\bibnamefont {Datta}},
  \bibinfo {author} {\bibfnamefont {R.}~\bibnamefont {Gavai}}, \ and\ \bibinfo
  {author} {\bibfnamefont {P.}~\bibnamefont {Majumdar}},\ }\Doi
  {10.1103/PhysRevD.85.014510} {\bibfield  {journal} {\bibinfo  {journal}
  {Phys. Rev.},\ }\textbf {\bibinfo {volume} {D85}},\ \bibinfo {pages} {014510}
  (\bibinfo {year} {2012})},\ \Eprint {http://arxiv.org/abs/1109.5738}
  {arXiv:1109.5738 [hep-lat]} \BibitemShut {NoStop}%
\bibitem [{\citenamefont {Francis}\ \emph
  {et~al.}(2015){\natexlab{a}}\citenamefont {Francis}, \citenamefont
  {Kaczmarek}, \citenamefont {Laine}, \citenamefont {Neuhaus},\ and\
  \citenamefont {Ohno}}]{Francis:2015daa}%
  \BibitemOpen
  \bibfield  {author} {\bibinfo {author} {\bibfnamefont {A.}~\bibnamefont
  {Francis}}, \bibinfo {author} {\bibfnamefont {O.}~\bibnamefont {Kaczmarek}},
  \bibinfo {author} {\bibfnamefont {M.}~\bibnamefont {Laine}}, \bibinfo
  {author} {\bibfnamefont {T.}~\bibnamefont {Neuhaus}}, \ and\ \bibinfo
  {author} {\bibfnamefont {H.}~\bibnamefont {Ohno}},\ }\Doi
  {10.1103/PhysRevD.92.116003} {\bibfield  {journal} {\bibinfo  {journal}
  {Phys. Rev.},\ }\textbf {\bibinfo {volume} {D92}},\ \bibinfo {pages} {116003}
  (\bibinfo {year} {2015}{\natexlab{a}})},\ \Eprint
  {http://arxiv.org/abs/1508.04543} {arXiv:1508.04543 [hep-lat]} \BibitemShut
  {NoStop}%
\bibitem [{\citenamefont {Altenkort}\ \emph {et~al.}(2019)\citenamefont
  {Altenkort}, \citenamefont {Kaczmarek}, \citenamefont {Mazur},\ and\
  \citenamefont {Shu}}]{Shu:2019twy}%
  \BibitemOpen
  \bibfield  {author} {\bibinfo {author} {\bibfnamefont {L.}~\bibnamefont
  {Altenkort}}, \bibinfo {author} {\bibfnamefont {O.}~\bibnamefont
  {Kaczmarek}}, \bibinfo {author} {\bibfnamefont {L.}~\bibnamefont {Mazur}}, \
  and\ \bibinfo {author} {\bibfnamefont {H.-T.}\ \bibnamefont {Shu}},\ }\Doi
  {10.22323/1.363.0204} {\bibfield  {journal} {\bibinfo  {journal} {PoS},\
  }\textbf {\bibinfo {volume} {LATTICE2019}},\ \bibinfo {pages} {204} (\bibinfo
  {year} {2019})},\ \Eprint {http://arxiv.org/abs/1912.11248} {arXiv:1912.11248
  [hep-lat]} \BibitemShut {NoStop}%
\bibitem [{\citenamefont {Lüscher}\ and\ \citenamefont
  {Weisz}(2001)}]{Luscher:2001up}%
  \BibitemOpen
  \bibfield  {author} {\bibinfo {author} {\bibfnamefont {M.}~\bibnamefont
  {Lüscher}}\ and\ \bibinfo {author} {\bibfnamefont {P.}~\bibnamefont
  {Weisz}},\ }\Doi {10.1088/1126-6708/2001/09/010} {\bibfield  {journal}
  {\bibinfo  {journal} {JHEP},\ }\textbf {\bibinfo {volume} {09}},\ \bibinfo
  {pages} {010} (\bibinfo {year} {2001})},\ \Eprint
  {http://arxiv.org/abs/hep-lat/0108014} {arXiv:hep-lat/0108014 [hep-lat]}
  \BibitemShut {NoStop}%
\bibitem [{\citenamefont {Bazavov}\ \emph
  {et~al.}(2018){\natexlab{a}}\citenamefont {Bazavov}, \citenamefont
  {Petreczky},\ and\ \citenamefont {Weber}}]{Bazavov:2017dsy}%
  \BibitemOpen
  \bibfield  {author} {\bibinfo {author} {\bibfnamefont {A.}~\bibnamefont
  {Bazavov}}, \bibinfo {author} {\bibfnamefont {P.}~\bibnamefont {Petreczky}},
  \ and\ \bibinfo {author} {\bibfnamefont {J.~H.}\ \bibnamefont {Weber}},\
  }\Doi {10.1103/PhysRevD.97.014510} {\bibfield  {journal} {\bibinfo  {journal}
  {Phys. Rev.},\ }\textbf {\bibinfo {volume} {D97}},\ \bibinfo {pages} {014510}
  (\bibinfo {year} {2018}{\natexlab{a}})},\ \Eprint
  {http://arxiv.org/abs/1710.05024} {arXiv:1710.05024 [hep-lat]} \BibitemShut
  {NoStop}%
\bibitem [{\citenamefont {Bazavov}\ \emph {et~al.}(2013)\citenamefont
  {Bazavov}, \citenamefont {Ding}, \citenamefont {Hegde}, \citenamefont
  {Karsch}, \citenamefont {Miao}, \citenamefont {Mukherjee}, \citenamefont
  {Petreczky}, \citenamefont {Schmidt},\ and\ \citenamefont
  {Velytsky}}]{Bazavov:2013uja}%
  \BibitemOpen
  \bibfield  {author} {\bibinfo {author} {\bibfnamefont {A.}~\bibnamefont
  {Bazavov}}, \bibinfo {author} {\bibfnamefont {H.~T.}\ \bibnamefont {Ding}},
  \bibinfo {author} {\bibfnamefont {P.}~\bibnamefont {Hegde}}, \bibinfo
  {author} {\bibfnamefont {F.}~\bibnamefont {Karsch}}, \bibinfo {author}
  {\bibfnamefont {C.}~\bibnamefont {Miao}}, \bibinfo {author} {\bibfnamefont
  {S.}~\bibnamefont {Mukherjee}}, \bibinfo {author} {\bibfnamefont
  {P.}~\bibnamefont {Petreczky}}, \bibinfo {author} {\bibfnamefont
  {C.}~\bibnamefont {Schmidt}}, \ and\ \bibinfo {author} {\bibfnamefont
  {A.}~\bibnamefont {Velytsky}},\ }\Doi {10.1103/PhysRevD.88.094021} {\bibfield
   {journal} {\bibinfo  {journal} {Phys. Rev.},\ }\textbf {\bibinfo {volume}
  {D88}},\ \bibinfo {pages} {094021} (\bibinfo {year} {2013})},\ \Eprint
  {http://arxiv.org/abs/1309.2317} {arXiv:1309.2317 [hep-lat]} \BibitemShut
  {NoStop}%
\bibitem [{\citenamefont {Ding}\ \emph {et~al.}(2015)\citenamefont {Ding},
  \citenamefont {Mukherjee}, \citenamefont {Ohno}, \citenamefont {Petreczky},\
  and\ \citenamefont {Schadler}}]{Ding:2015fca}%
  \BibitemOpen
  \bibfield  {author} {\bibinfo {author} {\bibfnamefont {H.~T.}\ \bibnamefont
  {Ding}}, \bibinfo {author} {\bibfnamefont {S.}~\bibnamefont {Mukherjee}},
  \bibinfo {author} {\bibfnamefont {H.}~\bibnamefont {Ohno}}, \bibinfo {author}
  {\bibfnamefont {P.}~\bibnamefont {Petreczky}}, \ and\ \bibinfo {author}
  {\bibfnamefont {H.~P.}\ \bibnamefont {Schadler}},\ }\Doi
  {10.1103/PhysRevD.92.074043} {\bibfield  {journal} {\bibinfo  {journal}
  {Phys. Rev.},\ }\textbf {\bibinfo {volume} {D92}},\ \bibinfo {pages} {074043}
  (\bibinfo {year} {2015})},\ \Eprint {http://arxiv.org/abs/1507.06637}
  {arXiv:1507.06637 [hep-lat]} \BibitemShut {NoStop}%
\bibitem [{\citenamefont {Bazavov}\ \emph {et~al.}(2016)\citenamefont
  {Bazavov}, \citenamefont {Brambilla}, \citenamefont {Ding}, \citenamefont
  {Petreczky}, \citenamefont {Schadler}, \citenamefont {Vairo},\ and\
  \citenamefont {Weber}}]{Bazavov:2016uvm}%
  \BibitemOpen
  \bibfield  {author} {\bibinfo {author} {\bibfnamefont {A.}~\bibnamefont
  {Bazavov}}, \bibinfo {author} {\bibfnamefont {N.}~\bibnamefont {Brambilla}},
  \bibinfo {author} {\bibfnamefont {H.~T.}\ \bibnamefont {Ding}}, \bibinfo
  {author} {\bibfnamefont {P.}~\bibnamefont {Petreczky}}, \bibinfo {author}
  {\bibfnamefont {H.~P.}\ \bibnamefont {Schadler}}, \bibinfo {author}
  {\bibfnamefont {A.}~\bibnamefont {Vairo}}, \ and\ \bibinfo {author}
  {\bibfnamefont {J.~H.}\ \bibnamefont {Weber}},\ }\Doi
  {10.1103/PhysRevD.93.114502} {\bibfield  {journal} {\bibinfo  {journal}
  {Phys. Rev.},\ }\textbf {\bibinfo {volume} {D93}},\ \bibinfo {pages} {114502}
  (\bibinfo {year} {2016})},\ \Eprint {http://arxiv.org/abs/1603.06637}
  {arXiv:1603.06637 [hep-lat]} \BibitemShut {NoStop}%
\bibitem [{\citenamefont {Bazavov}\ \emph
  {et~al.}(2018){\natexlab{b}}\citenamefont {Bazavov}, \citenamefont
  {Brambilla}, \citenamefont {Petreczky}, \citenamefont {Vairo},\ and\
  \citenamefont {Weber}}]{Bazavov:2018wmo}%
  \BibitemOpen
  \bibfield  {author} {\bibinfo {author} {\bibfnamefont {A.}~\bibnamefont
  {Bazavov}}, \bibinfo {author} {\bibfnamefont {N.}~\bibnamefont {Brambilla}},
  \bibinfo {author} {\bibfnamefont {P.}~\bibnamefont {Petreczky}}, \bibinfo
  {author} {\bibfnamefont {A.}~\bibnamefont {Vairo}}, \ and\ \bibinfo {author}
  {\bibfnamefont {J.~H.}\ \bibnamefont {Weber}} (\bibinfo {collaboration}
  {TUMQCD}),\ }\Doi {10.1103/PhysRevD.98.054511} {\bibfield  {journal}
  {\bibinfo  {journal} {Phys. Rev.},\ }\textbf {\bibinfo {volume} {D98}},\
  \bibinfo {pages} {054511} (\bibinfo {year} {2018}{\natexlab{b}})},\ \Eprint
  {http://arxiv.org/abs/1804.10600} {arXiv:1804.10600 [hep-lat]} \BibitemShut
  {NoStop}%
\bibitem [{\citenamefont {Bazavov}\ \emph {et~al.}(2019)\citenamefont {Bazavov}
  \emph {et~al.}}]{Bazavov:2019www}%
  \BibitemOpen
  \bibfield  {author} {\bibinfo {author} {\bibfnamefont {A.}~\bibnamefont
  {Bazavov}} \emph {et~al.},\ }\Doi {10.1103/PhysRevD.100.094510} {\bibfield
  {journal} {\bibinfo  {journal} {Phys. Rev. D},\ }\textbf {\bibinfo {volume}
  {100}},\ \bibinfo {pages} {094510} (\bibinfo {year} {2019})},\ \Eprint
  {http://arxiv.org/abs/1908.09552} {arXiv:1908.09552 [hep-lat]} \BibitemShut
  {NoStop}%
\bibitem [{\citenamefont {Burnier}\ \emph {et~al.}(2010)\citenamefont
  {Burnier}, \citenamefont {Laine}, \citenamefont {Langelage},\ and\
  \citenamefont {Mether}}]{Burnier:2010rp}%
  \BibitemOpen
  \bibfield  {author} {\bibinfo {author} {\bibfnamefont {Y.}~\bibnamefont
  {Burnier}}, \bibinfo {author} {\bibfnamefont {M.}~\bibnamefont {Laine}},
  \bibinfo {author} {\bibfnamefont {J.}~\bibnamefont {Langelage}}, \ and\
  \bibinfo {author} {\bibfnamefont {L.}~\bibnamefont {Mether}},\ }\Doi
  {10.1007/JHEP08(2010)094} {\bibfield  {journal} {\bibinfo  {journal} {JHEP},\
  }\textbf {\bibinfo {volume} {08}},\ \bibinfo {pages} {094} (\bibinfo {year}
  {2010})},\ \Eprint {http://arxiv.org/abs/1006.0867} {arXiv:1006.0867
  [hep-ph]} \BibitemShut {NoStop}%
\bibitem [{\citenamefont {Lüscher}(2010)}]{Luscher:2010iy}%
  \BibitemOpen
  \bibfield  {author} {\bibinfo {author} {\bibfnamefont {M.}~\bibnamefont
  {Lüscher}},\ }\Doi {10.1007/JHEP08(2010)071} {\bibfield  {journal} {\bibinfo
   {journal} {JHEP},\ }\textbf {\bibinfo {volume} {08}},\ \bibinfo {pages}
  {071} (\bibinfo {year} {2010})},\ \bibinfo {note} {[Erratum: JHEP 03, 092
  (2014)]},\ \Eprint {http://arxiv.org/abs/1006.4518} {arXiv:1006.4518
  [hep-lat]} \BibitemShut {NoStop}%
\bibitem [{\citenamefont {Francis}\ \emph
  {et~al.}(2015){\natexlab{b}}\citenamefont {Francis}, \citenamefont
  {Kaczmarek}, \citenamefont {Laine}, \citenamefont {Neuhaus},\ and\
  \citenamefont {Ohno}}]{Francis:2015lha}%
  \BibitemOpen
  \bibfield  {author} {\bibinfo {author} {\bibfnamefont {A.}~\bibnamefont
  {Francis}}, \bibinfo {author} {\bibfnamefont {O.}~\bibnamefont {Kaczmarek}},
  \bibinfo {author} {\bibfnamefont {M.}~\bibnamefont {Laine}}, \bibinfo
  {author} {\bibfnamefont {T.}~\bibnamefont {Neuhaus}}, \ and\ \bibinfo
  {author} {\bibfnamefont {H.}~\bibnamefont {Ohno}},\ }\Doi
  {10.1103/PhysRevD.91.096002} {\bibfield  {journal} {\bibinfo  {journal}
  {Phys. Rev.},\ }\textbf {\bibinfo {volume} {D91}},\ \bibinfo {pages} {096002}
  (\bibinfo {year} {2015}{\natexlab{b}})},\ \Eprint
  {http://arxiv.org/abs/1503.05652} {arXiv:1503.05652 [hep-lat]} \BibitemShut
  {NoStop}%
\bibitem [{\citenamefont {Christensen}\ and\ \citenamefont
  {Laine}(2016)}]{Christensen:2016wdo}%
  \BibitemOpen
  \bibfield  {author} {\bibinfo {author} {\bibfnamefont {C.}~\bibnamefont
  {Christensen}}\ and\ \bibinfo {author} {\bibfnamefont {M.}~\bibnamefont
  {Laine}},\ }\Doi {10.1016/j.physletb.2016.02.020} {\bibfield  {journal}
  {\bibinfo  {journal} {Phys. Lett.},\ }\textbf {\bibinfo {volume} {B755}},\
  \bibinfo {pages} {316} (\bibinfo {year} {2016})},\ \Eprint
  {http://arxiv.org/abs/1601.01573} {arXiv:1601.01573 [hep-lat]} \BibitemShut
  {NoStop}%
\bibitem [{\citenamefont {Sommer}(1994)}]{Sommer:1993ce}%
  \BibitemOpen
  \bibfield  {author} {\bibinfo {author} {\bibfnamefont {R.}~\bibnamefont
  {Sommer}},\ }\Doi {10.1016/0550-3213(94)90473-1} {\bibfield  {journal}
  {\bibinfo  {journal} {Nucl. Phys.},\ }\textbf {\bibinfo {volume} {B411}},\
  \bibinfo {pages} {839} (\bibinfo {year} {1994})},\ \Eprint
  {http://arxiv.org/abs/hep-lat/9310022} {arXiv:hep-lat/9310022 [hep-lat]}
  \BibitemShut {NoStop}%
\bibitem [{\citenamefont {Meyer}(2009)}]{Meyer:2009vj}%
  \BibitemOpen
  \bibfield  {author} {\bibinfo {author} {\bibfnamefont {H.~B.}\ \bibnamefont
  {Meyer}},\ }\Doi {10.1088/1126-6708/2009/06/077} {\bibfield  {journal}
  {\bibinfo  {journal} {JHEP},\ }\textbf {\bibinfo {volume} {06}},\ \bibinfo
  {pages} {077} (\bibinfo {year} {2009})},\ \Eprint
  {http://arxiv.org/abs/0904.1806} {arXiv:0904.1806 [hep-lat]} \BibitemShut
  {NoStop}%
\bibitem [{\citenamefont {Kapusta}\ and\ \citenamefont
  {Gale}(2011)}]{Kapusta:2006pm}%
  \BibitemOpen
  \bibfield  {author} {\bibinfo {author} {\bibfnamefont {J.}~\bibnamefont
  {Kapusta}}\ and\ \bibinfo {author} {\bibfnamefont {C.}~\bibnamefont {Gale}},\
  }\Doi {10.1017/CBO9780511535130} {\emph {\bibinfo {title}
  {{Finite-temperature field theory: Principles and applications}}}},\
  Cambridge Monographs on Mathematical Physics\ (\bibinfo  {publisher}
  {Cambridge University Press},\ \bibinfo {year} {2011})\ ISBN \bibinfo {isbn}
  {978-0-521-17322-3, 978-0-521-82082-0, 978-0-511-22280-1}\BibitemShut
  {NoStop}%
\bibitem [{\citenamefont {Tanabashi}\ \emph {et~al.}(2018)\citenamefont
  {Tanabashi} \emph {et~al.}}]{Tanabashi:2018oca}%
  \BibitemOpen
  \bibfield  {author} {\bibinfo {author} {\bibfnamefont {M.}~\bibnamefont
  {Tanabashi}} \emph {et~al.} (\bibinfo {collaboration} {Particle Data
  Group}),\ }\Doi {10.1103/PhysRevD.98.030001} {\bibfield  {journal} {\bibinfo
  {journal} {Phys. Rev.},\ }\textbf {\bibinfo {volume} {D98}},\ \bibinfo
  {pages} {030001} (\bibinfo {year} {2018})}\BibitemShut {NoStop}%
\bibitem [{\citenamefont {Kajantie}\ \emph {et~al.}(1997)\citenamefont
  {Kajantie}, \citenamefont {Laine}, \citenamefont {Rummukainen},\ and\
  \citenamefont {Shaposhnikov}}]{Kajantie:1997tt}%
  \BibitemOpen
  \bibfield  {author} {\bibinfo {author} {\bibfnamefont {K.}~\bibnamefont
  {Kajantie}}, \bibinfo {author} {\bibfnamefont {M.}~\bibnamefont {Laine}},
  \bibinfo {author} {\bibfnamefont {K.}~\bibnamefont {Rummukainen}}, \ and\
  \bibinfo {author} {\bibfnamefont {M.~E.}\ \bibnamefont {Shaposhnikov}},\
  }\Doi {10.1016/S0550-3213(97)00425-2} {\bibfield  {journal} {\bibinfo
  {journal} {Nucl. Phys.},\ }\textbf {\bibinfo {volume} {B503}},\ \bibinfo
  {pages} {357} (\bibinfo {year} {1997})},\ \Eprint
  {http://arxiv.org/abs/hep-ph/9704416} {arXiv:hep-ph/9704416 [hep-ph]}
  \BibitemShut {NoStop}%
\bibitem [{\citenamefont {Caron-Huot}\ and\ \citenamefont
  {Moore}(2008){\natexlab{b}}}]{CaronHuot:2007gq}%
  \BibitemOpen
  \bibfield  {author} {\bibinfo {author} {\bibfnamefont {S.}~\bibnamefont
  {Caron-Huot}}\ and\ \bibinfo {author} {\bibfnamefont {G.~D.}\ \bibnamefont
  {Moore}},\ }\Doi {10.1103/PhysRevLett.100.052301} {\bibfield  {journal}
  {\bibinfo  {journal} {Phys. Rev. Lett.},\ }\textbf {\bibinfo {volume}
  {100}},\ \bibinfo {pages} {052301} (\bibinfo {year} {2008}{\natexlab{b}})},\
  \Eprint {http://arxiv.org/abs/0708.4232} {arXiv:0708.4232 [hep-ph]}
  \BibitemShut {NoStop}%
\bibitem [{\citenamefont {Kovtun}\ \emph {et~al.}(2003)\citenamefont {Kovtun},
  \citenamefont {Son},\ and\ \citenamefont {Starinets}}]{Kovtun:2003wp}%
  \BibitemOpen
  \bibfield  {author} {\bibinfo {author} {\bibfnamefont {P.}~\bibnamefont
  {Kovtun}}, \bibinfo {author} {\bibfnamefont {D.~T.}\ \bibnamefont {Son}}, \
  and\ \bibinfo {author} {\bibfnamefont {A.~O.}\ \bibnamefont {Starinets}},\
  }\Doi {10.1088/1126-6708/2003/10/064} {\bibfield  {journal} {\bibinfo
  {journal} {JHEP},\ }\textbf {\bibinfo {volume} {10}},\ \bibinfo {pages} {064}
  (\bibinfo {year} {2003})},\ \Eprint {http://arxiv.org/abs/hep-th/0309213}
  {arXiv:hep-th/0309213 [hep-th]} \BibitemShut {NoStop}%
\bibitem [{\citenamefont {Andreev}(2018)}]{Andreev:2017bvr}%
  \BibitemOpen
  \bibfield  {author} {\bibinfo {author} {\bibfnamefont {O.}~\bibnamefont
  {Andreev}},\ }\Doi {10.1142/S0217732318500414} {\bibfield  {journal}
  {\bibinfo  {journal} {Mod. Phys. Lett. A},\ }\textbf {\bibinfo {volume}
  {33}},\ \bibinfo {pages} {1850041} (\bibinfo {year} {2018})},\ \Eprint
  {http://arxiv.org/abs/1707.05045} {arXiv:1707.05045 [hep-ph]} \BibitemShut
  {NoStop}%
\bibitem [{\citenamefont {Tolos}\ and\ \citenamefont
  {Torres-Rincon}(2013)}]{Tolos:2013kva}%
  \BibitemOpen
  \bibfield  {author} {\bibinfo {author} {\bibfnamefont {L.}~\bibnamefont
  {Tolos}}\ and\ \bibinfo {author} {\bibfnamefont {J.~M.}\ \bibnamefont
  {Torres-Rincon}},\ }\Doi {10.1103/PhysRevD.88.074019} {\bibfield  {journal}
  {\bibinfo  {journal} {Phys. Rev. D},\ }\textbf {\bibinfo {volume} {88}},\
  \bibinfo {pages} {074019} (\bibinfo {year} {2013})},\ \Eprint
  {http://arxiv.org/abs/1306.5426} {arXiv:1306.5426 [hep-ph]} \BibitemShut
  {NoStop}%
\bibitem [{\citenamefont {Torres-Rincon}\ \emph {et~al.}(2014)\citenamefont
  {Torres-Rincon}, \citenamefont {Tolos},\ and\ \citenamefont
  {Romanets}}]{Torres-Rincon:2014ffa}%
  \BibitemOpen
  \bibfield  {author} {\bibinfo {author} {\bibfnamefont {J.~M.}\ \bibnamefont
  {Torres-Rincon}}, \bibinfo {author} {\bibfnamefont {L.}~\bibnamefont
  {Tolos}}, \ and\ \bibinfo {author} {\bibfnamefont {O.}~\bibnamefont
  {Romanets}},\ }\Doi {10.1103/PhysRevD.89.074042} {\bibfield  {journal}
  {\bibinfo  {journal} {Phys. Rev. D},\ }\textbf {\bibinfo {volume} {89}},\
  \bibinfo {pages} {074042} (\bibinfo {year} {2014})},\ \Eprint
  {http://arxiv.org/abs/1403.1371} {arXiv:1403.1371 [hep-ph]} \BibitemShut
  {NoStop}%
\bibitem [{\citenamefont {Abreu}\ \emph {et~al.}(2011)\citenamefont {Abreu},
  \citenamefont {Cabrera}, \citenamefont {Llanes-Estrada},\ and\ \citenamefont
  {Torres-Rincon}}]{Abreu:2011ic}%
  \BibitemOpen
  \bibfield  {author} {\bibinfo {author} {\bibfnamefont {L.~M.}\ \bibnamefont
  {Abreu}}, \bibinfo {author} {\bibfnamefont {D.}~\bibnamefont {Cabrera}},
  \bibinfo {author} {\bibfnamefont {F.~J.}\ \bibnamefont {Llanes-Estrada}}, \
  and\ \bibinfo {author} {\bibfnamefont {J.~M.}\ \bibnamefont
  {Torres-Rincon}},\ }\Doi {10.1016/j.aop.2011.06.006} {\bibfield  {journal}
  {\bibinfo  {journal} {Annals Phys.},\ }\textbf {\bibinfo {volume} {326}},\
  \bibinfo {pages} {2737} (\bibinfo {year} {2011})},\ \Eprint
  {http://arxiv.org/abs/1104.3815} {arXiv:1104.3815 [hep-ph]} \BibitemShut
  {NoStop}%
\bibitem [{\citenamefont {Acharya}\ \emph {et~al.}(2018)\citenamefont {Acharya}
  \emph {et~al.}}]{Acharya:2017qps}%
  \BibitemOpen
  \bibfield  {author} {\bibinfo {author} {\bibfnamefont {S.}~\bibnamefont
  {Acharya}} \emph {et~al.} (\bibinfo {collaboration} {ALICE}),\ }\Doi
  {10.1103/PhysRevLett.120.102301} {\bibfield  {journal} {\bibinfo  {journal}
  {Phys. Rev. Lett.},\ }\textbf {\bibinfo {volume} {120}},\ \bibinfo {pages}
  {102301} (\bibinfo {year} {2018})},\ \Eprint
  {http://arxiv.org/abs/1707.01005} {arXiv:1707.01005 [nucl-ex]} \BibitemShut
  {NoStop}%
\bibitem [{\citenamefont {Adamczyk}\ \emph {et~al.}(2017)\citenamefont
  {Adamczyk} \emph {et~al.}}]{Adamczyk:2017xur}%
  \BibitemOpen
  \bibfield  {author} {\bibinfo {author} {\bibfnamefont {L.}~\bibnamefont
  {Adamczyk}} \emph {et~al.} (\bibinfo {collaboration} {STAR}),\ }\Doi
  {10.1103/PhysRevLett.118.212301} {\bibfield  {journal} {\bibinfo  {journal}
  {Phys. Rev. Lett.},\ }\textbf {\bibinfo {volume} {118}},\ \bibinfo {pages}
  {212301} (\bibinfo {year} {2017})},\ \Eprint
  {http://arxiv.org/abs/1701.06060} {arXiv:1701.06060 [nucl-ex]} \BibitemShut
  {NoStop}%
\end{thebibliography}%
